\documentclass{aa}  

\usepackage{graphicx}
\usepackage{txfonts}
\usepackage{lipsum}
\usepackage{subcaption}         
\usepackage{lscape}             
\usepackage{placeins}           
                                
\usepackage{xcolor}

\usepackage[normalem]{ulem}

\begin{document}

\title{Magnetic fields in blazar jets derived from synchrotron self-absorption}
\subtitle{}

%
%
%

   \author{Hubing Xiao\inst{1}\corrauth{hubing.xiao@shnu.edu.cn}     
        \and Junhang Liu\inst{1}\email{1000589536@smail.shnu.edu.cn}
        \and Shaopu Wang\inst{2,3}\email{2112419029@e.gzhu.edu.cn}
        \and Jianghe Yang\inst{4}\email{yjianghe@163.com}
        \and Zhihao Ouyang\inst{1}\email{zhihao.ouyang@outlook.com}
        \and Minfeng Gu\inst{5}\email{gumf@shao.ac.cn}
        \and Shaohua Zhang\inst{1}\email{zhangshaohua@shnu.edu.cn}
        \and Zhijian Luo\inst{1}\email{zjluo@shnu.edu.cn}
        \and Junhui Fan\inst{2,3,6,7}\email{fjh@gzhu.edu.cn}
        }

   \institute{Shanghai Key Lab for Astrophysics, Shanghai Normal University, Shanghai 200234, People’s Republic of China
   \and 
   Center for Astrophysics, Guangzhou University, Guangzhou 510006, People’s Republic of China
   \and
   Astronomy Science and Technology Research Laboratory of Department of Education of Guangdong Province, Guangzhou 510006, People’s Republic of China
   \and
   Department of Physics and Electronics Science, Hunan University of Arts and Science, Changde, 415000, People’s Republic of China
   \and
   Shanghai Astronomical Observatory, Chinese Academy of Sciences, Shanghai 200030, People’s Republic of China
   \and
   Great Bay Brand Center of the National Astronomical Data Center, Guangzhou 510006, People’s Republic of China
   \and
   Key Laboratory for Astronomical Observation and Technology of Guangzhou, Guangzhou 510006, People’s Republic of China
   }

   \date{\today}

 
  \abstract
   {Magnetic fields are believed to play a crucial role in the launching, acceleration, and evolution of blazar jets. 
   However, direct measurements of the magnetic field strength in the radio-emitting regions of blazar jets remain limited.}
   {We aim to estimate the magnetic field strengths in the synchrotron self-absorption (SSA) regions of blazar jets and investigate their connections with the properties of the jet and central engine.} 
   {We compiled multifrequency radio data for 143 blazars and fitted their radio spectra with a combined power-law and SSA model to derive the turnover frequency and turnover flux density. 
   The magnetic field strengths were then calculated by combining the SSA parameters with the very long baseline interferometry (VLBI) measured angular sizes and Doppler factors compiled from the literature.}
   {The SSA-derived magnetic fields are systematically lower than those obtained from broadband spectral energy distribution modeling, suggesting that the two methods probe different regions of the jet. 
   We further investigated the correlations between the SSA magnetic field strength and the apparent knot speed, $\gamma$-ray luminosity, black hole mass, and accretion disk luminosity.
   Only a weak correlation is found between the SSA magnetic field strength and $\gamma$-ray luminosity, while the best-fit relations with the other investigated parameters also exhibit negative slopes, suggesting a common tendency that is qualitatively consistent with magnetic energy being progressively converted into bulk kinetic energy during jet propagation.
   }
   {Our work presents a large statistical studies of SSA-derived magnetic fields in blazars and provides new observational constraints on the magnetic evolution of relativistic jets.}

   \keywords{active galactic nuclei --
                blazar --
                jet
               }

   \maketitle
   \nolinenumbers

\section{Introduction} \label{sec:intro}

The study of magnetic field strength is crucial to many areas of astrophysics. 
Although the nature of cosmic magnetic fields remains unknown, their strength is constrained to lie within the range of $10^{-16} \,{\rm G} \lesssim B \lesssim 10^{-9} \,{\rm G}$, based on studies of TeV emission cascades in blazars \citep{Neronov2010Sci328} and perturbations in the cosmic microwave background \citep{Planck2016A&A594}. 
The average magnetic field strength of the Milky Way, which plays an important role in shaping its spatial structure, is estimated to be $\sim 4\,{\rm \mu G}$, corresponding approximately to energy equipartition between the magnetic field and small-scale turbulent motions \citep{Dolginov1988PhR162}. 
Direct measurements by the Voyager 2 spacecraft yielded a magnetic field strength of $4-5\,{\rm \mu G}$, consistent with this estimate \citep{Opher2009Natur462}.

Active galactic nuclei (AGNs) are among the most persistent and powerful objects in the Universe, emitting radiation across the entire electromagnetic spectrum. 
An AGN is generally powered by a central supermassive black hole (SMBH) surrounded by an accretion disk. 
Active galactic nuclei are commonly classified as jetted or non-jetted systems depending on whether they launch relativistic jets \citep{Padovani2017NatAs1}. 
In AGN jets, magnetic fields play a pivotal role in jet formation, collimation, and particle acceleration. 
Magnetic fields anchored in the vicinity of the black hole or the accretion disk are believed to extract rotational energy and power relativistic jets through the Blandford--Znajek and Blandford--Payne mechanisms \citep{Blandford1977, Blandford1982}. 
Blazars constitute a class of jetted AGNs whose relativistic jets are oriented close to our line of sight and are therefore strongly affected by relativistic beaming. 
They are generally classified into flat-spectrum radio quasars (FSRQs) and BL Lacertae (BL Lacs) objects. 
Consequently, measuring the magnetic field strength in blazar jets is essential for understanding the formation, acceleration, and evolution of relativistic jets.

Several approaches have been developed to estimate the magnetic field in blazar jets, including the following:
(1) Broadband spectral energy distribution (SED) modeling.
Blazar SEDs generally exhibit a characteristic two-hump structure. 
The low-energy hump is generally attributed to synchrotron radiation from relativistic electrons, while the high-energy hump is commonly interpreted as arising from inverse Compton (IC) scattering of electrons or hadronic interactions. 
Within leptonic and hadronic frameworks, the magnetic field strength ($B$) in the emission region can be estimated by modeling the broadband SED \citep[e.g.,][]{Ghisellini2009MNRAS397, Chen2018, Xue2021ApJ906, Ouyang2025ApJ980, Xiao2025ApJ991, Xiao2026ApJ1005}.
(2) Model-based synchrotron luminosity.
\citet{Lee2016ApJ826} employed a standard jet model that assumes a steady, conical jet with a small opening angle \citep{Blandford1979ApJ232}. 
In this model, the magnetic field strength decreases with increasing distance ($d$) from the jet apex approximately as $d^{-1}$, while the electron energy density and magnetic field energy density are assumed to remain close to equipartition. 
Under these assumptions, the magnetic field strength can be constrained using Eq. (6) of \citet{Lee2016ApJ826}. 
A similar approach was adopted by \citet{Fan2023ApJS}, who constrained both the magnetic field strength and electron power for the largest blazar sample to date using synchrotron radiation equations within a one-zone leptonic model.
(3) The core-shift effect.
The unresolved radio core of a relativistic jet corresponds approximately to the region where the optical depth to synchrotron self-absorption (SSA) is of order unity. 
Because the jet becomes optically thick at lower frequencies and optically thinner at higher frequencies, the apparent core position shifts toward the central black hole with increasing observing frequency \citep{Blandford1979ApJ232, Konigl1981ApJ243, Lobanov1998A&A330}.
The magnetic field strength can then be estimated from the measured core shift under additional assumptions or model constraints \citep[e.g.,][]{Zdziarski2015MNRAS451, Agarwal2017MNRAS469, Nokhrina2024MNRAS528}.
(4) Synchrotron self-absorption.
The flat radio spectra commonly observed in blazars can be explained by the superposition of synchrotron self-absorbed spectra from multiple emitting regions with different turnover frequencies, as demonstrated, for example, by the radio spectrum of the quasar NRAO 140 \citep{Marscher1983ApJ264, Marscher1988ApJ334}. 
The magnetic field strength in an SSA region can be estimated by modeling the synchrotron self-absorbed spectrum and combining the resulting SSA parameters with additional constraints such as the source size and Doppler factor \citep[e.g.,][]{Rani2013, Lee2017ApJ841, Hodgson2017A&A597, Algaba2018ApJ859, Kim2022MNRAS510}.
(5) Faraday rotation.
Polarization observations provide another means of probing magnetic fields in blazar jets. 
When linearly polarized synchrotron emission propagates through magnetized plasma, its polarization position angle undergoes Faraday rotation. 
The magnitude of this rotation, characterized by the rotation measure (RM), depends on the thermal electron density and the line-of-sight component of the magnetic field. 
Thus, under appropriate assumptions about the plasma density and geometry, Faraday rotation measurements can be used to constrain the magnetic field strength and investigate the magnetic-field structure of blazar jets \citep{Gabuzda2017Galax5, Park2022Galax10}.
This method has been widely applied to investigate the magnetic-field structure of blazar jets \citep[e.g.,][]{Asada2002PASJ54, OSullivan2009MNRAS393, Gabuzda2015MNRAS450, Hovatta2019AA, Toscano2025A&A698}.

In this work, we applied the SSA method to radio data compiled from the NASA/IPAC Extragalactic Database (NED)\footnote{\url{https://ned.ipac.caltech.edu/}} to estimate the magnetic field strengths in the SSA regions of blazars and investigate the evolution of magnetic fields along relativistic jets.
This paper is organized as follows. 
Section 2 describes the data and methodology. 
Section 3 presents the results.
Section 4 presents the discussion. 
Section 5 summarizes our main conclusions.

\section{The SSA method and data acquisition}
\subsection{SSA spectral model}

The broadband radio spectra of blazars are generally interpreted as the superposition of emission from a long-lived underlying jet and one or more compact synchrotron self-absorbed components associated with newly ejected disturbances, such as shocks propagating along the jet \citep[e.g.,][]{Marscher1985ApJ298, Turler2000AA361}. 
Following this picture, we modeled the observed radio spectrum as the sum of two components: a quiescent power-law continuum representing the underlying jet emission and a homogeneous SSA component describing the perturbed emission region.

The underlying jet emission is approximated described by a power-law spectrum,
\begin{equation}
S_{\nu}=c_{\rm u}\nu^{\alpha_{\rm u}},
\end{equation}
where $\alpha_{\rm u}$ is the spectral index and $c_{\rm u}$ is the normalization factor.
The emission from the compact disturbed region is described by the homogeneous SSA model \citep[e.g.,][]{Turler2000AA361,Fromm2011AA531,Rani2013}:
\begin{equation}
S_{\nu}=S_{\rm m}
\left(\frac{\nu}{\nu_{\rm m}}\right)^{\alpha_{\rm t}}
\frac{1-\exp\left[-\tau_{\rm m}\left(\nu/\nu_{\rm m}\right)^{\alpha_0-\alpha_{\rm t}}\right]}
{1-\exp(-\tau_{\rm m})},
\end{equation}
where $\tau_{\rm m}\approx \frac{3}{2} \left(\sqrt{1-\frac{8\alpha_0}{3\alpha_{\rm t}}}-1\right)$
is the optical depth at the turnover frequency $\nu_{\rm m}$. 
Here, $\nu_{\rm m}$ and $S_{\rm m}$ denote the turnover frequency and turnover flux density, respectively, while $\alpha_{\rm t}$ and $\alpha_{0}$ are the spectral indices in the optically thick and optically thin regimes ($S_{\nu}\propto\nu^{\alpha}$). 
The total radio spectrum is therefore expressed as the sum of the underlying power-law continuum and the SSA component.

\subsection{Radio data collection and sample construction}
In the present work, we collected multifrequency radio-band data from \citet{Yang2022ApJS}, whose original observations were compiled from the NED. 
The collected data span frequencies from tens of megahertz (MHz) to tens of gigahertz (GHz).
The radio observations available in NED were obtained at different epochs; thus, the data used for the radio spectral analysis are not simultaneous, and intrinsic source variability may introduce additional uncertainties into the reconstructed radio spectra. 
We therefore constructed an overall broadband radio spectrum by combining the available measurements across different frequencies and adopted the average flux density for frequencies with multi-epoch observations. 
Although the resulting spectra do not represent strictly simultaneous emission states, the overall broadband spectral shapes are expected to provide a statistical representation of the underlying, time-averaged SSA properties of the sources.

To apply the SSA method and estimate the magnetic field strengths in blazar jets, we further refined the sample by requiring each source to satisfy the following criteria:
(1) exhibit a significant convex or turnover spectral feature at radio frequencies, indicative of SSA;
(2) have spectral measurements at no fewer than six distinct frequencies.
Applying these criteria, we compiled a final sample of 143 blazars, including 19 BL Lacs, 113 FSRQs, and 11 blazar candidates of uncertain type (BCUs). 
The source classifications were originally obtained from the \textit{Fermi} Large Area Telescope (LAT) Fourth Source Catalog (4FGL) \citep{Abdollahi2022ApJS260} and updated following \citet{Xiao2022ApJ_2}, who reported type transitions for 52 blazars.

\section{Results}
\subsection{The radio spectral fitting}
The radio spectra were fitted using the combined model described in previous section.
To minimize parameter degeneracy in the radio spectral modeling, we adopted a two-step fitting procedure rather than fitting all model parameters simultaneously.

The underlying continuum component was first determined independently. 
Since the SSA turnover can significantly affect the observed spectral shape, only spectral measurements expected to be minimally influenced by the SSA component were used in this stage. 
These measurements were selected manually based on the spectral distribution, including points located at the lowest and highest observed frequencies, the minimum-flux measurements within specified frequency intervals
considered representative of the continuum emission. 
The selected data were then fitted with a simple power-law model to determine the continuum parameters.

After establishing the continuum baseline, the best-fit power-law parameters were fixed and incorporated into the total spectral model. 
The SSA component was subsequently fitted to the full radio spectrum, with only the SSA-related parameters allowed to vary.
This approach effectively treats the power-law continuum as a fixed background contribution and enables the turnover characteristics to be constrained primarily by the SSA component.
During the fitting, $\alpha_{\rm t}$ was set to vary in the range from 1 to 3 and $\alpha_{\rm 0}$ was set to vary from $-$1.5 to $-$0.5.

All model fitting was performed using the nonlinear least-squares algorithm implemented in the SciPy package (`curve\_fit'), with observational uncertainties adopted as fitting weights. 
For measurements lacking reported uncertainties, the uncertainty was replaced by the mean uncertainty of the corresponding source to avoid singular weights during the fitting procedure.

\begin{figure*}[htbp]
\centering
\includegraphics[scale=0.6]{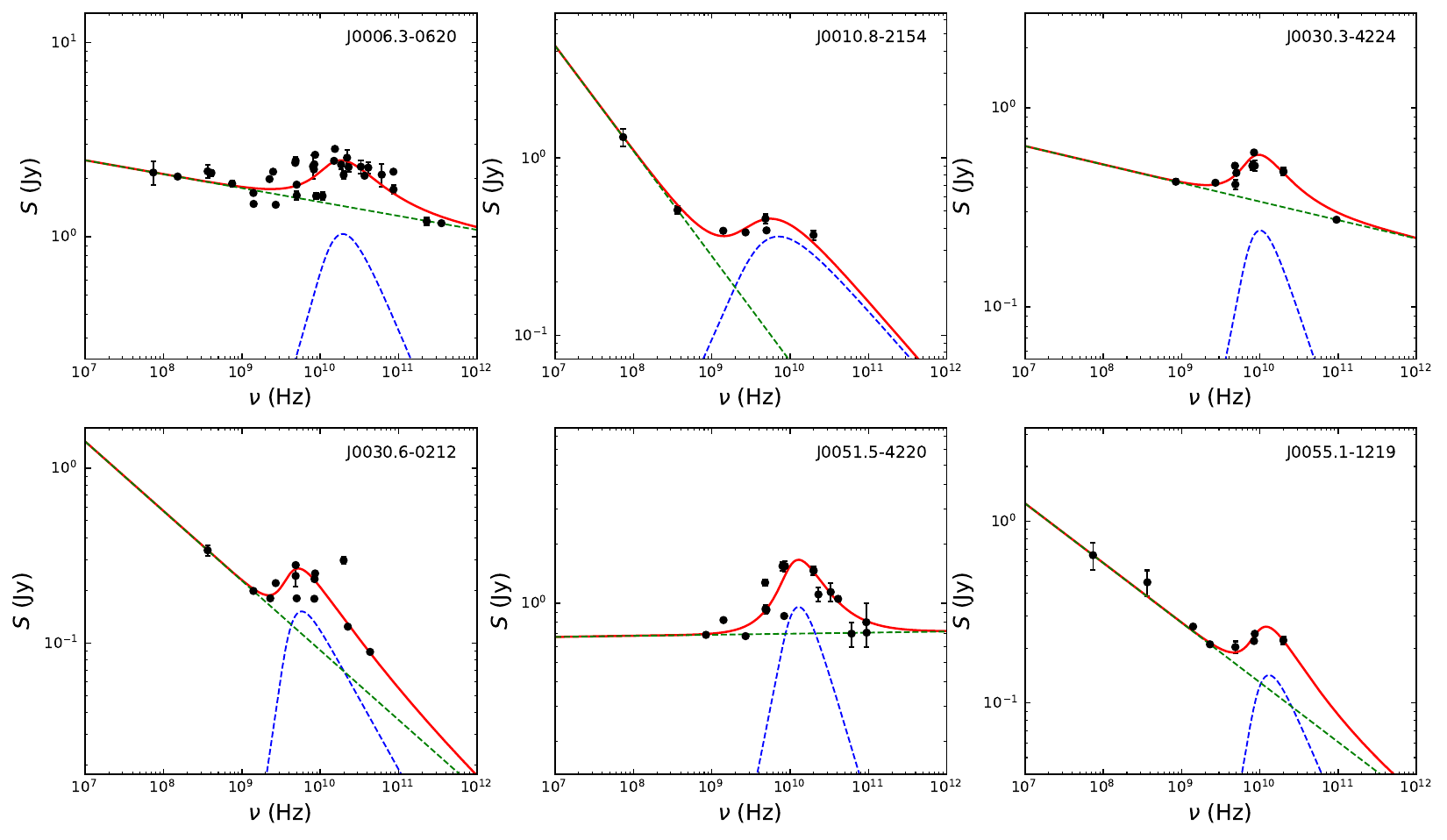}
\caption{Radio spectral fit results for the SSA region of the 143 blazars.
The black points indicate the observed data with errors. 
The unperturbed spectrum (dashed green line), the perturbed SSA spectrum (dashed blue line), and the total spectrum (solid red line) are also plotted.
The entire figure of plots is available in Sect. \ref{data}.}
\label{fig_ssa}
\end{figure*}

The quality of the fit was evaluated using the reduced chi-square statistic ($\chi_{\rm reduced}^{2}$): 
\begin{equation}
    \chi^{2}_{\rm reduced} = \frac{1}{N - N_{\rm p}}\sum^{N}_{i} \frac{(O_{i} - E_{i})^2}{\sigma^{2}_{i}},
\end{equation}
where $N_{\rm p}$ is the number of independently fitted parameters, and $N$ is the the number of data points. Here, $O_{i}$ and $E_{i}$ are the observed and model expected quantities, $\sigma_{i}$ is the uncertainty, while some flux measurements are reported with zero uncertainties. 
For these points, we assigned uncertainties equal to the mean uncertainty of the other valid data points for the same source. 
The final spectral decomposition consists of a power-law continuum component, an SSA component, and their combined model, which were subsequently used for the analysis of the radio spectral properties.
For a small number of sources with unsatisfactory automatic fitting results, the initial parameter estimates were manually adjusted and the fitting procedure was repeated to obtain physically reasonable solutions.
The fitting results are presented in Table \ref{tab_ssa} and Fig. \ref{fig_ssa}. The distribution of $\log S_{\rm m}$ and $\log \nu_{\rm m}$ are shown in the top and middle panel of Fig. \ref{fig_hist}.

\begin{table*}[htbp]
\centering
\caption{Best-fit parameters of the radio spectral models.}
\label{tab_ssa}
\begin{tabular}{lcccccccccc}
\hline
4FGL Name & J2000 Name & Class & z & $\log S_{\rm m}$ &	$\log \nu_{\rm m}$ & $\alpha_{\rm t}$ &	$\alpha_{\rm 0}$ & $\log c_{\rm u}$ & $\alpha_{\rm u}$ &$\chi^2_{\rm reduced}$	\\
 & & & & Jy & Hz & &  & & &  \\    
(1)	& (2) &	(3)	& (4) &	(5)	& (6) &	(7)	& (8) &	(9)	& (10) & (11) \\ 
\hline
J0006.3$-$0620 & J0006$-$0623 & B & 0.346676 & 0.01  & 10.28 & 1.41 & $-$0.96 & 0.89  & 0.07  & 22.4 \\
J0010.8$-$2154 & J0010$-$2157 & U &	         & $-$0.44 &  9.84 & 1.00 & $-$0.50 & 4.78  & 0.59  & 10.9 \\
J0030.3$-$4224 & J0030$-$4224 & F & 0.492    & $-$0.62 & 10.00 & 2.00 & $-$1.18 & 0.46  & 0.09  & 5.5 \\
J0030.6$-$0212 & J0030$-$0211 &	U & 0.5      & $-$0.82 & 9.76	 & 3.06 & $-$0.81	& 2.94  & 0.40  & 44.0 \\
J0051.5$-$4220 & J0051$-$4226 &	F &1.749	 & $-$0.02 & 10.10 & 2.00	& $-$1.34	& $-$0.21 & $-$0.01 & 13.4 \\
\hline
\end{tabular}
\tablefoot{
Column (1) 4FGL name;
column (2) J2000 name;
column (3) class "B" denotes BL Lac, "F" denotes FSRQ, and "U¨ denotes BCU;
column (4) redshift;
column (5) logarithmic flux density at the SSA peak, in units of jansky (Jy);
column (6) logarithmic frequency at the SSA peak, in units of hertz (Hz);
column (7) spectral index for the thick part of SSA spectrum;
column (8) spectral index for the thin part of SSA spectrum;
column (9) logarithmic normalization factor for the spectrum of the unperturbed component;
column (10) spectral index for the spectrum of the unperturbed component;
column (11) reduced $\chi^{2}$ value.
Only five items are displayed; the entire table is available in Sect. \ref{data}.
}
\end{table*}

\subsection{Estimating the magnetic field strength}
The magnetic field strength ($B_{\rm SSA}$) in the SSA region can be expressed as 
\begin{equation}
    B_{\rm SSA} = 10^{-5} b(\alpha_{0}) S_{\rm m}^{-2} d_{\rm m}^{4} \nu_{\rm m}^{5} \left( \frac{\delta}{1+z}\right)^{-1} \,\, {\rm [G]},
\label{eq_b}
\end{equation}
where $b(\alpha_{0})$ is a dimensionless parameter, $d_{\rm m}$ is the angular size of core at $\nu_{\rm m}$ in mas, $\nu_{\rm m}$ is the turnover frequency in GHz, $S_{\rm m}$ is the flux density at $\nu_{\rm m}$ in Jy, $z$ is the redshift, and the $\delta$ is the Doppler factor \citep{Marscher1983ApJ264, Kim2022MNRAS510}.
We employed the formulation of \citet{Kim2022MNRAS510}, in which the Doppler factor term, $\delta/(1+z)$, enters with a power of $-1$ rather than $+1$, as appropriate for a steady radio core.
This formulation was adopted because our radio spectra were constructed from archival, nonsimultaneous observations collected over extended periods, intended to characterize the time-averaged properties of the synchrotron self-absorbed jet rather than those of an individual transient state.
$b(\alpha_{0})$ is associated with the optically thin spectral index ($\alpha_0$) and was computed following the formalism in the Appendix A of \citet{Pushkarev2019MNRAS482}. 
In particular, we adopted their Eqs. (A6)–(A8) for the relevant physical constants, their Eq.(A11) for the optical depth, and their Eq. (A14) for the coefficient $b(\alpha_{0})$.
The angular size is estimated as $d_{\rm m} \simeq 1.8\,d_{\rm FWHM}$ \citep{Marscher1983ApJ264,Marscher1987ApJ319}, where $d_{\rm FWHM}$ is the Gaussian full width at half maximum (FWHM) measured by the very long baseline interferometry (VLBI).
This conversion accounts for the difference between the Gaussian FWHM and the angular diameter of an equivalent spherical source that produces a nearly identical visibility curve.
The value $d_{\rm FWHM}$ was obtained through by cross-matching our sample with the VLBI measurements compiled by \citet{Koryukova2022MNRAS515} and \citet{Hsu2023MNRAS525}.
For each source, the turnover frequency $\nu_{\rm m}$ derived from the radio spectral fitting was used as the reference frequency for the matching procedure.
The matching was performed in two steps. 
First, sources were identified through their source names. 
For each matched source, VLBI angular size measurements obtained at frequencies within $\pm$0.1 dex of the fitted $\nu_{\rm m}$ were preferentially selected. 
When multiple measurements satisfied this criterion, all corresponding entries were retained.
If no measurement was available within $\pm$0.1 dex of $\nu_{\rm m}$, a secondary search was performed using a wider frequency interval of $\pm$0.3 dex. 
Sources without VLBI observations within this expanded frequency range were excluded from the angular-size analysis.
Then, we averaged the these measurements of angular size for each sources and list them in Table \ref{tab_ddb}.
This procedure ensures that the adopted angular sizes are measured at frequencies as close as possible to the spectral turnover frequency, thereby minimizing the effects of frequency-dependent source structure and providing a more reliable estimate of the source size associated with the SSA-emitting region.
Ideally, the Doppler factor measured at or close to the turnover frequency ($\nu_{\rm m}$) should be adopted when calculating the SSA magnetic field strength $B_{\rm SSA}$. 
However, Doppler factors are not available at $\nu_{\rm m}$ for every source, and the available measurements are generally obtained at different observing frequencies. 
Moreover, the turnover frequencies $\nu_{\rm m}$ span a relatively broad range across our sample. 
Therefore, we divided the sources into five frequency bins according to their $\nu_{\rm m}$ values and assigned a representative Doppler factor to each bin. 
Sources with $\nu_{\rm m}$ falling within the same frequency bin were assigned the same representative Doppler factor. 
For each frequency bin, the representative Doppler factor was calculated as the average of the Doppler factors measured at frequencies within the corresponding band for blazars in the literature.
Specifically, sources with $\nu_{\rm m} \leq 10$ GHz were assigned $\langle \delta^{10} \rangle = 13.89 \pm 2.17$, obtained from the average Doppler factors of sources observed below 10 GHz in \citet{Liodakis2017MNRAS466}, in which they estimated Doppler factors at ten frequencies between 2.64 and 142.33 GHz for 58 $\gamma$-ray bright blazars.
Sources with $10 < \nu_{\rm m} \leq 20$ GHz were assigned $\langle \delta^{15} \rangle = 14.35 \pm 5.77$ from the 15 GHz sample of \citet{Liodakis2018}, in which they provided a homogeneous catalog of 15 GHz variability Doppler factors for 1029 sources. 
For sources with $20 < \nu_{\rm m} \leq 30$ GHz and $30 < \nu_{\rm m} \leq 40$ GHz, we adopted $\langle \delta^{22} \rangle = 13.36 \pm 8.09$ and $\langle \delta^{37} \rangle = 10.66 \pm 7.32$, respectively, based on the 22 and 37 GHz measurements reported by \citet{Hovatta2009}, in which they derived variability Doppler factors from long-term total flux density monitoring at 22 and 37 GHz for 87 AGNs. 
Finally, for sources with $40 < \nu_{\rm m} \leq 50$ GHz, we adopt $\langle \delta^{43} \rangle = 15.51 \pm 3.39$, derived from the 43 GHz VLBA sample of \citet{Jorstad2017ApJ846}, in which they derived Doppler factors at 43 GHz for 36 blazars using VLBA observations.
This frequency-dependent assignment minimizes the mismatch between the observing frequency at which the Doppler factor is measured and the turnover frequency of the source. 
This assignment thereby provides a more physically motivated estimate of relativistic beaming effects than adopting a single characteristic Doppler factor for the entire sample.
We employed the average redshift $\langle z \rangle = 1.053$ in our sample for the 11 sources without available redshift values, including 8 BCUs and 3 BL Lacs.

\begin{figure}[htbp]
\centering
\includegraphics[scale=0.8]{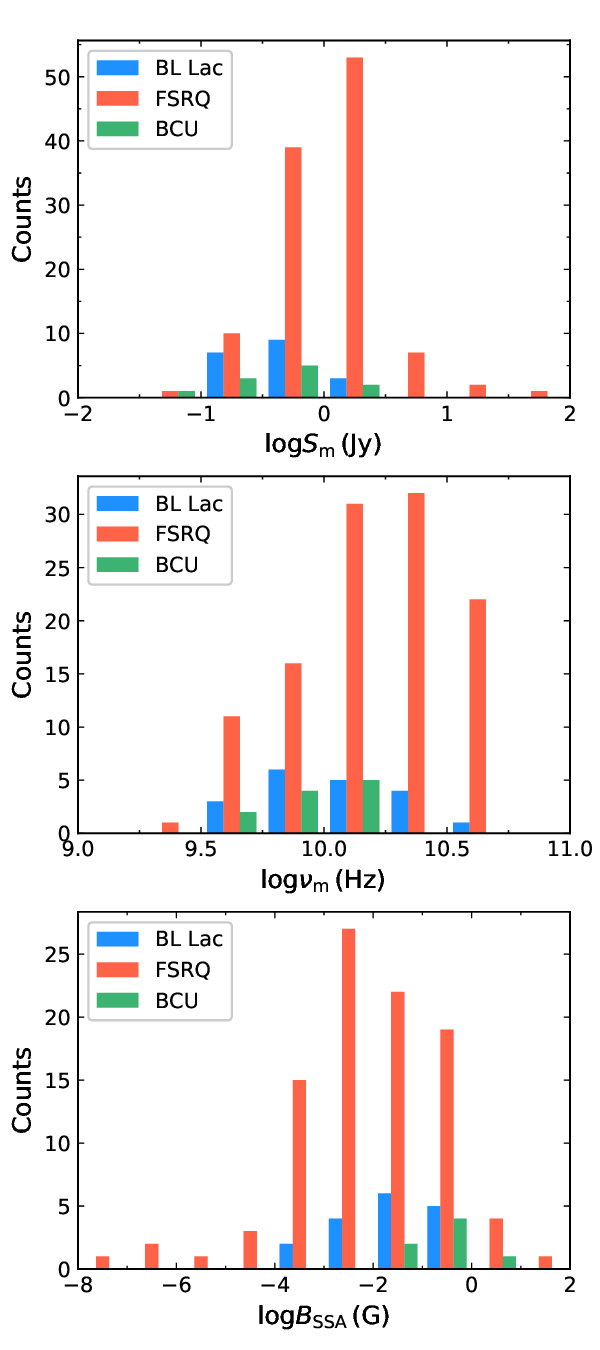}
\caption{Distribution of $\log S_{\rm m}$, $\log \nu_{\rm m}$, and $\log B_{\rm SSA}$.
The blue bar stands for BL Lacs, the red bar stands for FSRQs, and the green bar stands for BCUs.}
\label{fig_hist}
\end{figure}

In the end, we managed to calculate $B_{\rm SSA}$ for 119 sources with available $\langle d_{\rm FWHM} \rangle$ and list them together with $d_{\rm FWHM}$ and $\delta$ in Table \ref{tab_ddb}. 
A distribution of $\log B_{\rm SSA}$ is shown in the bottom panel of Fig. \ref{fig_hist}.
We notice four sources with $B_{\rm SSA} < 10^{-5} \, {\rm G}$, including three with $B_{\rm SSA} < 10^{-6} \, {\rm G}$. 
Such a small value is mainly due to relatively small $d_{\rm m}$.
And we should be careful with these small magnetic field strength, because it is significantly smaller than that obtained in literature \citep[e.g.,][]{Kim2022MNRAS510}.
The magnetic field strength of these four sources are not be employed in the following discussions and analyzing.

\begin{table*}[htbp]
\centering
\caption{VLBI core sizes and Doppler factors used to estimate the SSA magnetic field strengths.}
\label{tab_ddb}
\begin{tabular}{lccccccccc}
\hline
4FGL Name & J2000 Name & $\nu$ range & $\langle d_{\rm FWHM}\rangle$ &	match type	& Ref & $\delta$ & $\delta$ band & Ref & $\log B_{\rm SSA}$	\\
 & & GHz & mas & & &  & GHz & & Gs \\    
(1)	& (2) &	(3)	& (4) &	(5) & (6) & (7) & (8) & (9) & (10) \\ 
\hline
J0006.3$-$0620 & J0006$-$0623 & 15.29 - 15.39 & $0.198\pm0.003$ & within\_0.1\_dex & K22 & 14.35 &    15 & L18 & $-0.93 \pm 0.01$  \\
J0010.8$-$2154 & J0010$-$2157 &  8.34 - 8.68  & $0.261\pm0.079$ & within\_0.1\_dex & K22 & 13.89 & $<$10 & L17 & $-1.70 \pm 0.13$  \\
J0030.3$-$4224 & J0030$-$4224 &  8.36 - 8.68  & $0.090\pm0.078$	& within\_0.3\_dex & K22 & 13.89 & $<$10 & L17 & $-2.35 \pm 0.38$  \\
J0030.6$-$0212 & J0030$-$0211 &  8.65 - 8.68  & $0.132\pm0.024$	& within\_0.1\_dex & K22 & 13.89 & $<$10 & L17 & $-2.53 \pm 0.08$  \\
J0051.5$-$4220 & J0051$-$4226 &  8.64 - 8.68  & $0.156\pm0.237$	& within\_0.3\_dex & K22 & 14.35 &    15 & L18 & $-1.78 \pm 0.66$  \\
\hline
\end{tabular}
\tablefoot{
Column (1) 4FGL name;
column (2) J2000 name;
column (3) frequency of collected $d_{\rm m}$;
column (4) the average value of collected $d_{\rm m}$ in the $\nu$ range, in units of mas;
column (5) the criteria of collecting $d_{\rm m}$;
column (6) the reference of $d_{\rm m}$;
column (7) Doppler factor;
column (8) Doppler factor in each band;
column (9) the Doppler factor reference.
column (10) the logarithmic magnetic field strength.
Only five items are displayed; the entire table is available in Sect. \ref{data}.}
\end{table*}

\section{Discussion} \label{sec:dis}
\subsection{Systematic uncertainties and limitations}
During the fitting process, 10 (7.0\%) and 18 sources (12.6\%) reached the imposed boundaries of $\alpha_{\rm t}$ and $\alpha_{\rm 0}$, respectively, in the automatic fitting procedure. 
In addition, 90 sources (62.9\%) required manual adjustment of the initial parameter estimates to obtain physically reasonable and convergent solutions.
Such fitting instabilities are most likely a consequence of the nonsimultaneous radio observations.
Since the flux density measurements compiled from NED were obtained at different epochs, the reconstructed spectra do not represent the instantaneous SEDs of the sources.
Given the strong variability commonly observed in blazars, particularly at high radio frequencies, variability-induced distortions may shift the derived turnover frequency ($\nu_{\rm m}$) and peak flux density ($S_{\rm m}$).
Besides, the fitting instabilities also originate from the use of both single-dish and interferometric-array flux densities rather than only VLBI measurements. 
While this choice avoids complications caused by the highly heterogeneous angular resolutions of different VLBI observations, it also means that the measured flux densities may contain contributions from extended jet emission in addition to the compact core. 
Such contamination is expected to be more significant at lower frequencies and may affect the determination of the optically thin spectral slope and turnover parameters.
The SSA magnetic field depends strongly on these parameters, particularly on $\nu_{\rm m}$, and nonsimultaneous observations can introduce systematic uncertainties into the derived magnetic field strengths.

During the SSA magnetic field estimation using Eq.~\ref{eq_b}, the average redshift of the sample was used for 11 sources lacking redshift measurements, and six of them made it to $B_{\rm SSA}$ estimation.
This should bring uncertainties to the results. However, they are not significant.
Another source of systematic uncertainty arises from the Doppler factors. 
Since frequency-specific Doppler factors are unavailable for most objects, we adopted frequency-dependent average values derived from literature. 
Individual sources may differ substantially from these average Doppler factors, leading to uncertainties in the magnetic field estimates of individual objects. 
Although uncertainties in the Doppler factor propagate directly into the magnetic field estimates, their impact on the statistical properties of the sample is expected to be limited because the Doppler factor enters Eq.~\ref{eq_b} as an inverse linear factor.
An important source of uncertainty arises from the determination of $d_{\rm FWHM}$ and that of $d_{\rm m}$.
The angular sizes adopted in this work were obtained by matching the fitted turnover frequencies with VLBI measurements available within a narrow frequency interval around $\nu_{\rm m}$.
Although this procedure minimizes the effects of frequency-dependent source structure, the turnover frequencies and VLBI angular sizes are generally not measured simultaneously. 
Given the strong variability of blazars, temporal changes in the jet structure may lead to inconsistencies between the spectral turnover properties and the corresponding VLBI sizes. 
In addition, the VLBI measurements were compiled from different observing programs and may be subject to systematic differences in angular resolution and model-fitting procedures. 
Since the SSA magnetic field depends strongly on $d_{\rm m}$, uncertainties in the adopted angular sizes can propagate significantly into the derived magnetic field strengths. 
Nevertheless, by selecting measurements obtained at frequencies as close as possible to $\nu_{\rm m}$, the potential bias introduced by frequency-dependent core-size effects is substantially reduced, making the adopted $d_{\rm m}$ values suitable for statistical studies of large samples.

The ideal approach would be to employ quasi-simultaneous multifrequency radio observations, high-resolution VLBI measurements that isolate the compact core emission, and source-specific Doppler factors. 
However, such data are currently available only for a limited number of objects and would substantially reduce the sample size.
Despite these limitations, several factors support the reliability of the present statistical analysis. 
First, the sample contains 143 blazars spanning a broad range of spectral properties, reducing the influence of extreme behavior in individual sources. 
Second, the primary objective of this work is to investigate the global characteristics and distributions of magnetic field strengths rather than to obtain highly precise measurements for individual objects. 

The average turnover frequency, peak flux density, and SSA magnetic field strength for the sources in our sample are
$\langle \log \nu_{\rm m} \rangle = 10.15 \pm 0.30 \, {\rm Hz}$, 
$\langle \log S_{\rm m} \rangle = -0.05 \pm 0.44 \, {\rm Jy}$, and
$\langle \log B_{\rm SSA} \rangle = -1.81 \pm 0.16 \, {\rm G}$.
They are generally consistent with the values reported in previous SSA study of individual blazar jets \citep[e,g.,][]{Rani2013, Kim2022MNRAS510, Xu2026arXiv}.
Therefore, although the magnetic field estimates for individual sources should be interpreted with caution, the adopted methodology is adequate for exploring the statistical properties of the sample and for investigating the overall trends between magnetic field strength and other jet parameters.

\subsection{The magnetic field of the synchrotron self-absorption region}
In the framework of the conical jet model proposed by \citet{Blandford1979ApJ232} and further developed by \citet{Konigl1981ApJ243}, the observed VLBI radio core is generally interpreted as the surface where the optical depth reaches $\tau_{\nu} \approx 1$.
Owing to SSA, the apparent position of the radio core is frequency dependent, with higher-frequency observations probing regions progressively closer to the central engine. 
This frequency-dependent core shift has been both theoretically predicted and observationally confirmed in numerous AGN jets \citep{Lobanov1998A&A330, Pushkarev2012A&A545}. 
Consequently, the SSA turnover frequency is expected to be associated with the compact emitting region near the VLBI core, where the transition from optically thick to optically thin synchrotron emission occurs.

For blazars, whose radio emission is dominated by relativistically beamed jets viewed at small angles to the line of sight, the turnover region is generally believed to originate within the inner parsec-scale jet.
Observations indicate that radio cores detected at frequencies ranging from a few GHz to several tens of GHz typically correspond to distances of sub-parsec to several parsecs from the central black hole \citep{Lobanov1998A&A330, Pushkarev2012A&A545, Boccardi2017ARA&A}.

We calculated the average value of magnetic field strength
$\langle \log B_{\rm SSA}^{\rm F} = -1.90 \pm 0.15 \rangle$ and $\langle \log B_{\rm SSA}^{\rm B} = -1.78 \pm 0.25 \rangle$ for FSRQs and BL Lacs, respectively.
The result of Anderson-Darling test gives a statistic of $-$0.78 that is close to 0 and suggests the distribution of $\log B$ have no significant difference between FSRQs and BL Lacs.

Based on one-zone leptonic SED modeling, several studies have reported that FSRQs tend to possess stronger magnetic fields than BL Lac objects.
For example, \citet{Zhang2012ApJ752} found that the typical magnetic field strengths in FSRQs are systematically larger than those in BL Lac objects, which was interpreted as a consequence of their more powerful accretion systems and stronger jet powers. 
Similar conclusions have been obtained in our previous work \citet{Fan2023ApJS}, where the stronger magnetic fields in FSRQs contribute to enhanced radiative cooling and lower synchrotron peak frequencies.

The magnetic fields derived from the SSA method provide an independent probe of the physical conditions in the compact radio-emitting regions and are largely free from the model degeneracies inherent in broadband SED fitting.
Therefore, a comparison between the SSA magnetic fields of FSRQs and BL Lac objects can provide valuable constraints on the magnetization properties of blazar jets and may help clarify the role of magnetic fields in shaping the observed differences between the two blazar subclasses.
However, our result for $\log B_{\rm SSA}$ cannot constrain the magnetization properties of blazar subclasses.
Such a result may arise from an unbalanced sample size, because we have five times more FSRQs than BL Lacs in our sample.

\subsection{The evolution of blazar jet}
\begin{figure}[htbp]
\centering
\includegraphics[scale=0.8]{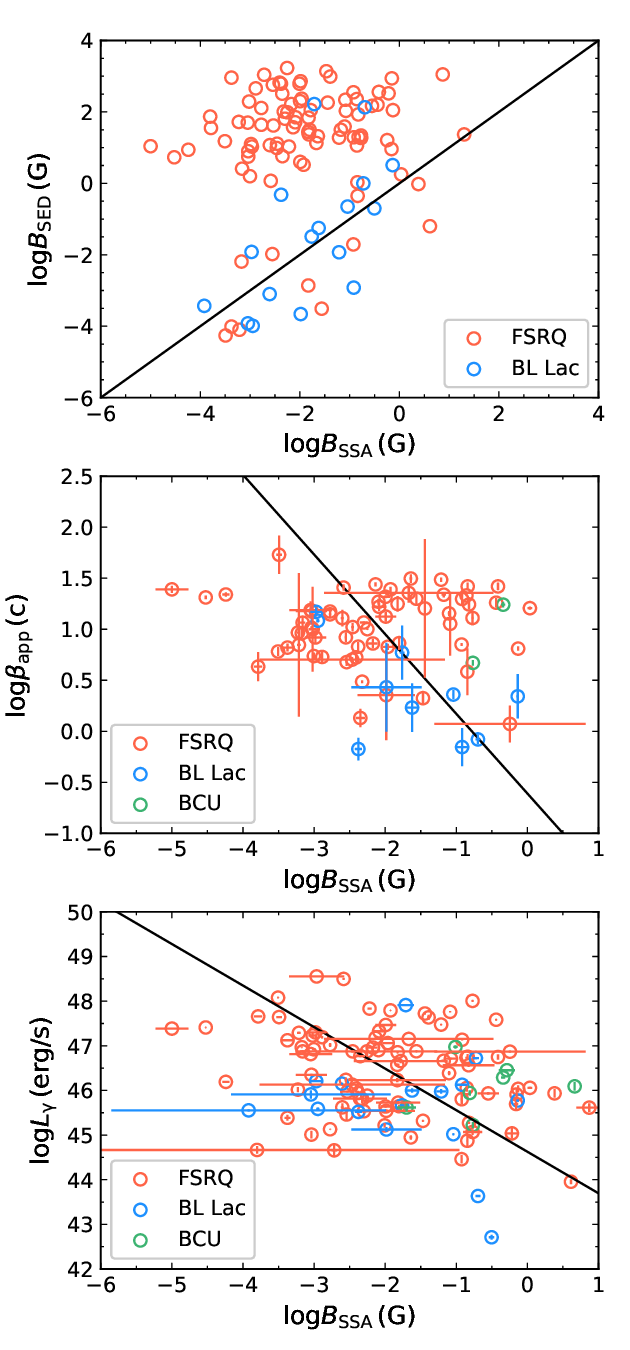}
\caption{Correlation between $\log B_{\rm SSA}$ from this work, $\log B_{\rm SED}$ from \citet{Fan2023ApJS}, apparent velocity ($\log \beta_{\rm app}$), and $\gamma$-ray luminosity ($\log L_{\gamma}$).
The blue circle stands for BL Lacs, the red circle stands for FSRQs, and the green circle stands for BCUs.}
\label{fig_jet_evo}
\end{figure}

We compare the magnetic field strengths derived in this work with those reported by \citet{Fan2023ApJS}. 
In their study, the magnetic fields ($B_{\rm SED}$) were estimated by applying synchrotron radiation theory to the low-energy component of the SEDs, where the synchrotron peak parameters were obtained from the parabolic SED fitting presented by \citet{Yang2022ApJS}. 
The comparison is shown in the top panel of Fig.~\ref{fig_jet_evo}. 
We find that the SSA magnetic fields derived in this work are systematically lower than the corresponding values of $B_{\rm SED}$.

This systematic difference likely reflects the fact that the two methods probe different emission regions within the jet. 
The magnetic fields derived from SED modeling characterize the compact high-energy emission region located within or around 0.1 pc from the central super massive black hole \citep{Fan2023ApJS}.
By contrast, the SSA method measures the magnetic field in the radio self-absorbed region associated with the spectral turnover, which is generally believed to be located farther downstream in the parsec-scale jet. 
If the magnetic field decreases with distance from the central engine, as predicted by magnetically driven jet models, the lower magnetic field strengths derived from the SSA method are naturally expected.
Therefore, our results are qualitatively consistent with the scenario in which the jet is initially Poynting-flux-dominated and progressively converts magnetic energy into bulk kinetic energy during its propagation \citep{Blandford1977, Blandford1982, Vlahakis2004ApJ605, Komissarov2007MNRAS380}.

If magnetic energy is progressively converted into bulk kinetic energy during jet propagation, the magnetic field strength is expected to decrease as the jet accelerates. 
Since the apparent knot speed ($\beta_{\rm app}$) is commonly adopted as a proxy for the jet bulk motion, we further investigated the relation between $\log B_{\rm SSA}$ and $\log \beta_{\rm app}$.
We collected the maximum apparent knot speeds measured from multi-epoch VLBI observations for 83 sources in our sample from \citet{Xiao2019SCPMA, Xiao2022MNRAS}. 
The maximum value of $\beta_{\rm app}$ was adopted because it is generally considered to provide the best estimate of the bulk jet speed. 
The correlation between $\log \beta_{\rm app}$ and $\log B_{\rm SSA}$ is presented in the middle panel of Fig.~\ref{fig_jet_evo}. 
The ordinary least-squares bisector (OLS-Bisector) regression yields
$$\log \beta_{\rm app} = -(0.78\pm0.19)\log B_{\rm SSA} - (0.61\pm0.39),$$
with a Pearson correlation coefficient of $r=-0.00$ and a chance probability of $p=0.97$.
These results indicate no statistically significant correlation between the magnetic field strength and the apparent knot speed.

And if magnetic energy is ultimately converted into particle energy and radiation, sources with weaker magnetic fields may also be expected to exhibit stronger high-energy emission.
We therefore investigated the relation between $\log B_{\rm SSA}$ and the $\gamma$-ray luminosity ($\log L_{\gamma}$), where the latter is widely regarded as a good tracer of the nonthermal bolometric output of blazars \citep{Ghisellini2014,Xiong2014,Zhang2020,Xiao2022ApJ_1}. 
The isotropic $\gamma$-ray luminosity is calculated as
\begin{equation}
L_{\rm \gamma} = 4 \pi d_{\rm L}^2(1+z)^{\alpha_{\rm ph}-2} F,
\label{eq_L_g}
\end{equation}
where $d_{\rm L} = (1+z) \cdot \frac{c}{H_{\rm 0}} \cdot \int_{\rm 1}^{\rm 1+z} \frac{1}{\sqrt{\Omega_{\rm M}x^3+1-\Omega_{\rm M}}}dx$, $z$ is redshift, $(1+z)^{\alpha_{\rm ph}-2}$ represents a $K$-correction, $\alpha_{\rm ph}$ is the $\gamma$-ray photon index, and $F$ is the $\gamma$-ray flux in units of $\rm erg \cdot cm^{\rm -2} \cdot s^{\rm -1}$ obtained through ``Flux1000'' (the integral photon flux from 1 to 100 GeV) from 4FGL.
The OLS-Bisector linear regression gives a result of 
$$\log L_{\gamma} = -(0.93\pm0.04)\log B_{\rm SSA} + (44.63\pm0.07),$$
with a Pearson correlation coefficient of $r=-0.17$ and a chance probability of $p=0.04$. 
Therefore, a weak correlation is found between the SSA magnetic field strength and the $\gamma$-ray luminosity.

Although no statistically significant correlations are found, both OLS-Bisector regressions yield negative slopes, indicating a tendency for sources with weaker SSA magnetic fields to exhibit larger apparent knot speeds and higher $\gamma$-ray luminosities. 
Such tendencies are qualitatively consistent with the magnetically driven jet scenario, in which magnetic energy is gradually converted into bulk kinetic energy and subsequently dissipated into particle acceleration and radiation during jet propagation.
Nevertheless, these intrinsic trends are expected to be substantially weakened by relativistic beaming and projection effects. 
In particular, the apparent knot speed depends on both the bulk Lorentz factor and the viewing angle, while the observed $\gamma$-ray luminosity is strongly Doppler-boosted. 
Moreover, the SSA magnetic field derived in this work characterizes the local magnetic field in the radio self-absorbed region rather than the total magnetic energy carried by the jet. 
These factors are likely responsible for the large scatter in the observed relations and the absence of statistically significant correlations.

\subsection{The central engine of blazar jet}
\begin{figure}[htbp]
\centering
\includegraphics[scale=0.8]{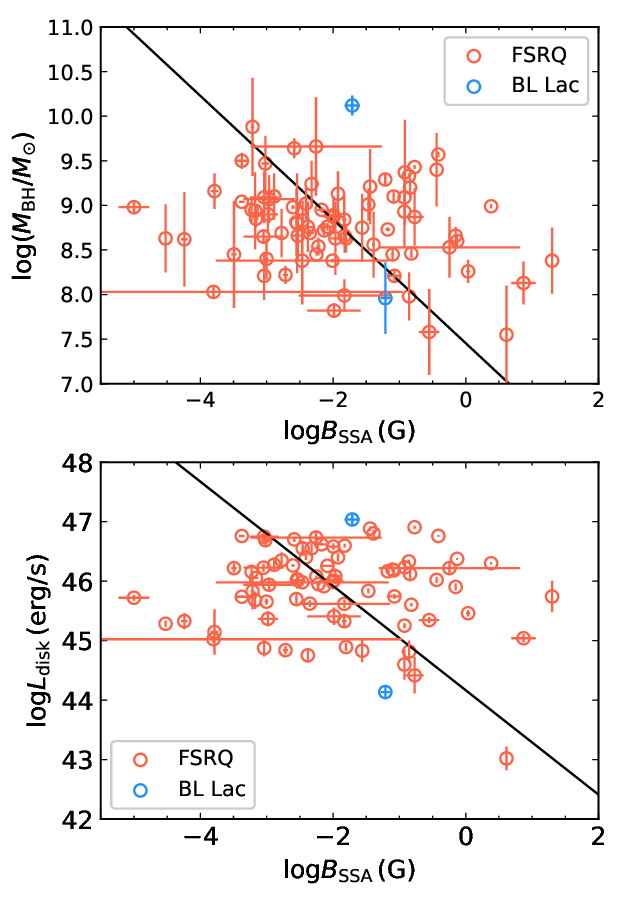}
\caption{Correlation between $\log B_{\rm SSA}$ from this work, black hole mass $\log M_{\rm BH}$, and accretion disk luminosity $\log L_{\rm disk}$.
The blue circle stands for BL Lacs, the red circle stands for FSRQs, and the green circle stands for BCUs.}
\label{fig_jet_BD}
\end{figure}

We compiled the black hole mass ($\log \frac{M_{\rm BH}}{M_{\odot}}$) and accretion disk luminosity ($\log L_{\rm disk}$) from \citet{Paliya2021} and \citet{Xiao2022ApJ_1}, in which these two quantities were estimated through spectral studies in the optical band.
In total, there are 97 sources in our sample with available $\log \frac{M_{\rm BH}}{M_{\odot}}$ and $\log L_{\rm disk}$.
The correlation between $\log \frac{M_{\rm BH}}{M_{\odot}}$ and $\log B_{\rm SSA}$ is presented in the top panel of Fig~\ref{fig_jet_BD}. 
The OLS-Bisector regression yields
$$\log \frac{M_{\rm BH}}{M_{\odot}} = -(0.69\pm0.14)\log B_{\rm SSA} + (7.45\pm0.27),$$
with a Pearson correlation coefficient of $r=-0.18$ and a chance probability of $p=0.1$.
The correlation between $\log L_{\rm disk}$ and $\log B_{\rm SSA}$ is presented in the bottom panel of Fig.~\ref{fig_jet_BD}. 
The OLS-Bisector regression yields
$$\log L_{\rm disk} = -(0.88\pm0.11)\log B_{\rm SSA} + (44.16\pm0.21),$$
with a Pearson correlation coefficient of $r=-0.12$ and a chance probability of $p=0.31$.

Both the $\log (M_{\rm BH}/M_{\odot})$--$\log B_{\rm SSA}$ and $\log L_{\rm disk}$--$\log B_{\rm SSA}$ relations exhibit negative slopes, and neither correlation is statistically significant.
This result is not unexpected, because the SSA magnetic field derived in this work characterizes the radio self-absorbed region in the inner parsec-scale jet rather than the immediate vicinity of the central engine.

As the jet propagates outward, the magnetic field strength is expected to decrease owing to jet expansion and the gradual conversion of magnetic energy into bulk kinetic energy through relativistic magnetohydrodynamics (MHD) acceleration \citep{Vlahakis2004ApJ605, Komissarov2007MNRAS380}.
In addition, magnetic reconnection, shocks, turbulence, and interactions with the ambient medium may further modify the magnetic field configuration along the jet. These evolutionary processes are likely to introduce substantial source-to-source scatter, thereby weakening any intrinsic dependence of the parsec-scale magnetic field on the central black hole properties.

Nevertheless, the best-fit regressions for both relations have negative slopes, suggesting a weak tendency for sources with larger black hole masses or higher disk luminosities to possess lower SSA magnetic fields.
Such a common tendency, which aligns with the tendency discussed in the previous section, is qualitatively consistent with the picture of magnetically driven jet evolution. In this scenario, magnetic energy is gradually converted into bulk kinetic energy and particle energy as the jet propagates downstream. 
Consequently, sources with more efficient jet acceleration or energy dissipation may exhibit lower magnetic field strengths in the SSA-emitting region.

Given the relatively low correlation coefficients and the lack of statistical significance, this tendency should be interpreted with caution. 
A more robust assessment requires larger samples with quasi-simultaneous multifrequency radio observations, more accurate measurements of the SSA-emitting region, and source-specific Doppler factors to reduce the systematic uncertainties in the magnetic field estimates.

\section{Conclusion}

In this work, we investigated the magnetic field strengths in the inner jets of blazars using the SSA method. 
By fitting the radio spectra with a combined power-law and SSA model, we derived the turnover frequencies and flux densities for a sample of 143 blazars. 
Combining these quantities with VLBI angular-size measurements and Doppler factors compiled from the literature, we estimated the magnetic field strengths in the SSA-emitting regions and explored their connections with jet and central-engine properties. 
Our main conclusions are summarized as follows.

(1) We successfully reproduced the radio spectra of 143 blazars with a combined power-law and SSA model and obtained the corresponding SSA turnover parameters. 

(2) The magnetic fields derived from the SSA method are systematically lower than those obtained from broadband SED modeling by \citet{Fan2023ApJS}. 
Since the two methods are expected to probe different emission regions, this result suggests that the magnetic field strength decreases from the compact high-energy emission region toward the downstream radio core.
Such a trend is qualitatively consistent with the theoretical picture of magnetically driven jets, in which magnetic energy is progressively converted into bulk kinetic energy during jet propagation.

(3) We investigated the relations between $\log B_{\rm SSA}$, $\log \beta_{\rm app}$, $\log L_{\gamma}$, $\log M_{\rm BH}$, and $\log L_{\rm disk}$. 
However, $\log B_{\rm SSA}$ only shows a weak correlation with $\log L_{\gamma}$, while the best-fit relations exhibit negative slopes for all the four quantities.
These common tendencies are qualitatively consistent with the expected evolution of magnetized relativistic jets, but the intrinsic relations are likely weakened by relativistic beaming, projection effects, source-to-source diversity, and the fact that the SSA magnetic field characterizes the parsec-scale radio core rather than the jet-launching region.

This work presents the largest statistical investigation of SSA-derived magnetic fields in blazars to date.
By combining radio spectral fitting with VLBI measurements, we provide a practical approach to estimating magnetic fields in the radio core and offer new observational constraints on the magnetic evolution of relativistic jets. 
Future quasi-simultaneous broadband radio observations, together with higher-resolution VLBI imaging and more accurate source-specific Doppler factors, will substantially improve the reliability of SSA magnetic field measurements and enable a more direct test of jet acceleration and magnetic energy dissipation models.

\section{Data availability}
\label{data}
The complete tables of the radio spectral fitting parameters (full Table \ref{tab_ssa}) and derived magnetic field strengths (full Table \ref{tab_ddb}), together with the full set of radio spectral fitting figures (full Fig. \ref{fig_ssa}), are publicly available on Zenodo at https://doi.org/10.5281/zenodo.23383536.

\begin{acknowledgements} 
H.B.X. acknowledges the support from National Natural Science Foundation of China (NSFC, grant No. 12203034), the Shanghai Science and Technology Fund (grant No. 22YF1431500), the China Manned Space Project (grant No. CMS-CSST-2025-A07), and the Shanghai Municipal Education Commission regarding artificial intelligence empowered research.
M.F.G. acknowledges the support from National Science Foundation of China (NSFC, grant No. 12473019), the Shanghai Pilot Program for Basic Research-Chinese Academy of Science, Shanghai Branch (grant No. JCYJ-SHFY-2021-013), the National SKA Program of China (grant No. 2022SKA0120102), and the China Manned Space Project (grant No. CMS-CSST-2025-A07).
Z.J.L. acknowledges the support from the National Natural Science Foundation of China (Grant No. 12573009) and the scientific research grants from the China Manned Space Project with Grand No. CMS-CSST-2025-A07 and CMS-CSST-2025-A05.
S.H.Z acknowledges the support from National Natural Science Foundation of China (NSFC, grant No. 12173026), the Program for Professor of Special Appointment (Eastern Scholar) at Shanghai Institutions of Higher Learning and the Shuguang Program (23SG39) of the Shanghai Education Development Foundation and Shanghai Municipal Education Commission.
J.H.F. acknowledges the support from National Natural Science Foundation of China (NSFC, grant No. 12433004), the National Key Research and Development Program of China (grant No. 2025YFA1614102), and the support from the Astrophysics Key Subjects of Guangdong Province and Guangzhou City.
\end{acknowledgements}

%

\bibliographystyle{aa}
\bibliography{lib_xiao}

@ARTICLE{Pushkarev2019MNRAS482,
       author = {{Pushkarev}, A.~B. and {Butuzova}, M.~S. and {Kovalev}, Y.~Y. and {Hovatta}, T.},
        title = "{Multifrequency study of the gamma-ray flaring BL Lacertae object PKS 2233-148 in 2009-2012}",
      journal = {\mnras},
         year = 2019,
        month = jan,
       volume = {482},
       number = {2},
        pages = {2336-2353},
          doi = {10.1093/mnras/sty2724},
archivePrefix = {arXiv},
       eprint = {1808.06138},
 primaryClass = {astro-ph.HE},
       adsurl = {https://ui.adsabs.harvard.edu/abs/2019MNRAS.482.2336P}
}

@ARTICLE{Marscher1987ApJ319,
       author = {{Marscher}, Alan P. and {Broderick}, John J. and {Padrielli}, Lucia and {Bartel}, Norbert and {Romney}, Jonathan D.},
        title = "{18 Centimeter VLBI Observations of the Quasar NRAO 140 during and after a Low-Frequency Outburst}",
      journal = {\apj},
         year = 1987,
        month = aug,
       volume = {319},
        pages = {456},
          doi = {10.1086/165470},
       adsurl = {https://ui.adsabs.harvard.edu/abs/1987ApJ...319..456M}
}

@ARTICLE{Nokhrina2024MNRAS528,
       author = {{Nokhrina}, E.~E. and {Pushkarev}, A.~B.},
        title = "{Core shift in parabolic accelerating jets}",
      journal = {\mnras},
         year = 2024,
        month = feb,
       volume = {528},
       number = {2},
        pages = {2523-2532},
          doi = {10.1093/mnras/stae179},
archivePrefix = {arXiv},
       eprint = {2401.06698},
 primaryClass = {astro-ph.HE},
       adsurl = {https://ui.adsabs.harvard.edu/abs/2024MNRAS.528.2523N}
}

@ARTICLE{Lee2017ApJ841,
       author = {{Lee}, Jee Won and {Lee}, Sang-Sung and {Hodgson}, Jeffrey A. and {Kim}, Dae-Won and {Algaba}, Juan-Carlos and {Kang}, Sincheol and {Kang}, Jiman and {Kim}, Sungsoo S.},
        title = "{Interferometric Monitoring of Gamma-Ray Bright AGNs: S5 0716+714}",
      journal = {\apj},
         year = 2017,
        month = jun,
       volume = {841},
       number = {2},
          eid = {119},
        pages = {119},
          doi = {10.3847/1538-4357/aa72f7},
archivePrefix = {arXiv},
       eprint = {1705.06835},
 primaryClass = {astro-ph.GA},
       adsurl = {https://ui.adsabs.harvard.edu/abs/2017ApJ...841..119L}
}

@ARTICLE{Hodgson2017A&A597,
       author = {{Hodgson}, J.~A. and {Krichbaum}, T.~P. and {Marscher}, A.~P. and {Jorstad}, S.~G. and {Rani}, B. and {Marti-Vidal}, I. and {Bach}, U. and {Sanchez}, S. and {Bremer}, M. and {Lindqvist}, M. and {Uunila}, M. and {Kallunki}, J. and {Vicente}, P. and {Fuhrmann}, L. and {Angelakis}, E. and {Karamanavis}, V. and {Myserlis}, I. and {Nestoras}, I. and {Chidiac}, C. and {Sievers}, A. and {Gurwell}, M. and {Zensus}, J.~A.},
        title = "{Location of {\ensuremath{\gamma}}-ray emission and magnetic field strengths in OJ 287}",
      journal = {\aap},
         year = 2017,
        month = jan,
       volume = {597},
          eid = {A80},
        pages = {A80},
          doi = {10.1051/0004-6361/201526727},
archivePrefix = {arXiv},
       eprint = {1607.00725},
 primaryClass = {astro-ph.HE},
       adsurl = {https://ui.adsabs.harvard.edu/abs/2017A&A...597A..80H}
}

@ARTICLE{Algaba2018ApJ859,
       author = {{Algaba}, Juan-Carlos and {Lee}, Sang-Sung and {Rani}, Bindu and {Kim}, Dae-Won and {Kino}, Motoki and {Hodgson}, Jeffrey and {Zhao}, Guang-Yao and {Byun}, Do-Young and {Gurwell}, Mark and {Kang}, Sin-Cheol and {Kim}, Jae-Young and {Kim}, Jeong-Sook and {Kim}, Soon-Wook and {Park}, Jong-Ho and {Trippe}, Sascha and {Wajima}, Kiyoaki},
        title = "{Exploring the Variability of the Flat-spectrum Radio Source 1633+382. II. Physical Properties}",
      journal = {\apj},
         year = 2018,
        month = jun,
       volume = {859},
       number = {2},
          eid = {128},
        pages = {128},
          doi = {10.3847/1538-4357/aac2e7},
archivePrefix = {arXiv},
       eprint = {1805.02849},
 primaryClass = {astro-ph.HE},
       adsurl = {https://ui.adsabs.harvard.edu/abs/2018ApJ...859..128A}
}

@ARTICLE{Gabuzda2017Galax5,
       author = {{Gabuzda}, Denise},
        title = "{The Origin and Structure of the Magnetic Fields and Currents of AGN Jets}",
      journal = {Galaxies},
         year = 2017,
        month = feb,
       volume = {5},
       number = {1},
          eid = {11},
        pages = {11},
          doi = {10.3390/galaxies5010011},
       adsurl = {https://ui.adsabs.harvard.edu/abs/2017Galax...5...11G}
}

@ARTICLE{Park2022Galax10,
       author = {{Park}, Jongho and {Algaba}, Juan Carlos},
        title = "{Polarization Observations of AGN Jets: Past and Future}",
      journal = {Galaxies},
         year = 2022,
        month = oct,
       volume = {10},
       number = {5},
          eid = {102},
        pages = {102},
          doi = {10.3390/galaxies10050102},
archivePrefix = {arXiv},
       eprint = {2210.13819},
 primaryClass = {astro-ph.HE},
       adsurl = {https://ui.adsabs.harvard.edu/abs/2022Galax..10..102P}
}

@ARTICLE{Asada2002PASJ54,
       author = {{Asada}, Keiichi and {Inoue}, Makoto and {Uchida}, Yutaka and {Kameno}, Seiji and {Fujisawa}, Kenta and {Iguchi}, Satoru and {Mutoh}, Mutsumi},
        title = "{A Helical Magnetic Field in the Jet of 3C 273}",
      journal = {\pasj},
         year = 2002,
        month = jun,
       volume = {54},
        pages = {L39-L43},
          doi = {10.1093/pasj/54.3.L39},
archivePrefix = {arXiv},
       eprint = {astro-ph/0205497},
 primaryClass = {astro-ph},
       adsurl = {https://ui.adsabs.harvard.edu/abs/2002PASJ...54L..39A}
}

@ARTICLE{OSullivan2009MNRAS393,
       author = {{O'Sullivan}, S.~P. and {Gabuzda}, D.~C.},
        title = "{Three-dimensional magnetic field structure of six parsec-scale active galactic nuclei jets}",
      journal = {\mnras},
         year = 2009,
        month = feb,
       volume = {393},
       number = {2},
        pages = {429-456},
          doi = {10.1111/j.1365-2966.2008.14213.x},
archivePrefix = {arXiv},
       eprint = {0811.4426},
 primaryClass = {astro-ph},
       adsurl = {https://ui.adsabs.harvard.edu/abs/2009MNRAS.393..429O}
}

@ARTICLE{Toscano2025A&A698,
       author = {{Toscano}, Teresa and {Molina}, Sol N. and {G{\'o}mez}, Jos{\'e} L. and {Zeng}, Ai-Ling and {Dahale}, Rohan and {Cho}, Ilje and {Moriyama}, Kotaro and {Wielgus}, Maciek and {Fuentes}, Antonio and {Foschi}, Marianna and {R{\"o}der}, Jan and {Myserlis}, Ioannis and {Angelakis}, Emmanouil and {Zensus}, Anton},
        title = "{Helical magnetic field structure in 3C 273: A Faraday rotation analysis using multi-frequency polarimetric VLBA data}",
      journal = {\aap},
         year = 2025,
        month = jun,
       volume = {698},
          eid = {A210},
        pages = {A210},
          doi = {10.1051/0004-6361/202453542},
archivePrefix = {arXiv},
       eprint = {2412.18250},
 primaryClass = {astro-ph.HE},
       adsurl = {https://ui.adsabs.harvard.edu/abs/2025A&A...698A.210T}
}

@ARTICLE{Gabuzda2015MNRAS450,
       author = {{Gabuzda}, D.~C. and {Knuettel}, S. and {Reardon}, B.},
        title = "{Transverse Faraday-rotation gradients across the jets of 15 active galactic nuclei}",
      journal = {\mnras},
         year = 2015,
        month = jul,
       volume = {450},
       number = {3},
        pages = {2441-2450},
          doi = {10.1093/mnras/stv555},
archivePrefix = {arXiv},
       eprint = {1503.03411},
 primaryClass = {astro-ph.GA},
       adsurl = {https://ui.adsabs.harvard.edu/abs/2015MNRAS.450.2441G}
}

@ARTICLE{Agarwal2017MNRAS469,
       author = {{Agarwal}, A. and {Mohan}, P. and {Gupta}, Alok C. and {Mangalam}, A. and {Volvach}, A.~E. and {Aller}, M.~F. and {Aller}, H.~D. and {Gu}, M.~F. and {L{\"a}hteenm{\"a}ki}, A. and {Tornikoski}, M. and {Volvach}, L.~N.},
        title = "{Core shift effect in blazars}",
      journal = {\mnras},
         year = 2017,
        month = jul,
       volume = {469},
       number = {1},
        pages = {813-840},
          doi = {10.1093/mnras/stx847},
archivePrefix = {arXiv},
       eprint = {1704.03229},
 primaryClass = {astro-ph.GA},
       adsurl = {https://ui.adsabs.harvard.edu/abs/2017MNRAS.469..813A}
}

@ARTICLE{Xu2026arXiv,
       author = {{Xu}, Wancheng and {Cui}, Lang and {An}, Tao and {Frey}, S{\'a}ndor and {Wang}, Xin and {Liu}, Yuanqi and {Chang}, Ning and {Chen}, Liang and {Zhang}, Yingkang},
        title = "{Synchrotron Self-Absorption Spectral Modeling Reveals a Magnetically Driven Shock-in-Jet Scenario in Blazar 1156+295}",
      journal = {arXiv e-prints},
         year = 2026,
        month = mar,
          eid = {arXiv:2603.20701},
        pages = {arXiv:2603.20701},
          doi = {10.48550/arXiv.2603.20701},
archivePrefix = {arXiv},
       eprint = {2603.20701},
 primaryClass = {astro-ph.HE},
       adsurl = {https://ui.adsabs.harvard.edu/abs/2026arXiv260320701X}
}

@ARTICLE{Xiao2026ApJ1005,
       author = {{Xiao}, Hubing and {Ouyang}, Zhihao and {Yang}, Lili and {Zhu}, Jingtian and {Gu}, Minfeng and {Chen}, Liang and {Zhang}, Shaohua and {Luo}, Zhijian and {Fan}, Junhui},
        title = "{Chasing the Neutrino Blazar Candidates. II. SED Modeling with a Hadronic Model}",
      journal = {\apj},
         year = 2026,
        month = jul,
       volume = {1005},
       number = {1},
          eid = {30},
        pages = {30},
          doi = {10.3847/1538-4357/ae66e9},
archivePrefix = {arXiv},
       eprint = {2605.08775},
 primaryClass = {astro-ph.HE},
       adsurl = {https://ui.adsabs.harvard.edu/abs/2026ApJ..1005...30X}
}

@ARTICLE{Koryukova2022MNRAS515,
       author = {{Koryukova}, T.~A. and {Pushkarev}, A.~B. and {Plavin}, A.~V. and {Kovalev}, Y.~Y.},
        title = "{Tracing Milky Way scattering by compact extragalactic radio sources}",
      journal = {\mnras},
         year = 2022,
        month = sep,
       volume = {515},
       number = {2},
        pages = {1736-1750},
          doi = {10.1093/mnras/stac1898},
archivePrefix = {arXiv},
       eprint = {2201.04359},
 primaryClass = {astro-ph.GA},
       adsurl = {https://ui.adsabs.harvard.edu/abs/2022MNRAS.515.1736K}
}

@ARTICLE{Pushkarev2012A&A545,
       author = {{Pushkarev}, A.~B. and {Hovatta}, T. and {Kovalev}, Y.~Y. and {Lister}, M.~L. and {Lobanov}, A.~P. and {Savolainen}, T. and {Zensus}, J.~A.},
        title = "{MOJAVE: Monitoring of Jets in Active galactic nuclei with VLBA Experiments. IX. Nuclear opacity}",
      journal = {\aap},
         year = 2012,
        month = sep,
       volume = {545},
          eid = {A113},
        pages = {A113},
          doi = {10.1051/0004-6361/201219173},
archivePrefix = {arXiv},
       eprint = {1207.5457},
 primaryClass = {astro-ph.HE},
       adsurl = {https://ui.adsabs.harvard.edu/abs/2012A&A...545A.113P}
}

@ARTICLE{Boccardi2017ARA&A,
       author = {{Boccardi}, B. and {Krichbaum}, T.~P. and {Ros}, E. and {Zensus}, J.~A.},
        title = "{Radio observations of active galactic nuclei with mm-VLBI}",
      journal = {\text{The Astronomy and Astrophysics Review}},
         year = 2017,
        month = dec,
       volume = {25},
       number = {1},
          eid = {4},
        pages = {4},
          doi = {10.1007/s00159-017-0105-6},
archivePrefix = {arXiv},
       eprint = {1711.07548},
 primaryClass = {astro-ph.HE},
       adsurl = {https://ui.adsabs.harvard.edu/abs/2017A&ARv..25....4B}
}

@ARTICLE{Hsu2023MNRAS525,
       author = {{Hsu}, Po-Chih and {Koay}, Jun Yi and {Matsushita}, Satoki and {Hwang}, Chorng-Yuan and {Hovatta}, Talvikki and {Kiehlmann}, Sebastian and {Readhead}, Anthony and {Max-Moerbeck}, Walter and {Reeves}, Rodrigo},
        title = "{Milliarcsecond core size dependence of the radio variability of blazars}",
      journal = {\mnras},
         year = 2023,
        month = nov,
       volume = {525},
       number = {4},
        pages = {5105-5120},
          doi = {10.1093/mnras/stad2525},
archivePrefix = {arXiv},
       eprint = {2308.09626},
 primaryClass = {astro-ph.GA},
       adsurl = {https://ui.adsabs.harvard.edu/abs/2023MNRAS.525.5105H}
}

@ARTICLE{Liodakis2017MNRAS466,
       author = {{Liodakis}, I. and {Marchili}, N. and {Angelakis}, E. and {Fuhrmann}, L. and {Nestoras}, I. and {Myserlis}, I. and {Karamanavis}, V. and {Krichbaum}, T.~P. and {Sievers}, A. and {Ungerechts}, H. and {Zensus}, J.~A.},
        title = "{F-GAMMA: variability Doppler factors of blazars from multiwavelength monitoring}",
      journal = {\mnras},
         year = 2017,
        month = apr,
       volume = {466},
       number = {4},
        pages = {4625-4632},
          doi = {10.1093/mnras/stx002},
archivePrefix = {arXiv},
       eprint = {1701.01452},
 primaryClass = {astro-ph.HE},
       adsurl = {https://ui.adsabs.harvard.edu/abs/2017MNRAS.466.4625L}
}

@ARTICLE{Jorstad2017ApJ846,
       author = {{Jorstad}, Svetlana G. and {Marscher}, Alan P. and {Morozova}, Daria A. and {Troitsky}, Ivan S. and {Agudo}, Iv{\'a}n and {Casadio}, Carolina and {Foord}, Adi and {G{\'o}mez}, Jos{\'e} L. and {MacDonald}, Nicholas R. and {Molina}, Sol N. and {L{\"a}hteenm{\"a}ki}, Anne and {Tammi}, Joni and {Tornikoski}, Merja},
        title = "{Kinematics of Parsec-scale Jets of Gamma-Ray Blazars at 43 GHz within the VLBA-BU-BLAZAR Program}",
      journal = {\apj},
         year = 2017,
        month = sep,
       volume = {846},
       number = {2},
          eid = {98},
        pages = {98},
          doi = {10.3847/1538-4357/aa8407},
archivePrefix = {arXiv},
       eprint = {1711.03983},
 primaryClass = {astro-ph.GA},
       adsurl = {https://ui.adsabs.harvard.edu/abs/2017ApJ...846...98J}
}

@ARTICLE{Xue2021ApJ906,
       author = {{Xue}, Rui and {Liu}, Ruo-Yu and {Wang}, Ze-Rui and {Ding}, Nan and {Wang}, Xiang-Yu},
        title = "{A Two-zone Blazar Radiation Model for ``Orphan'' Neutrino Flares}",
      journal = {\apj},
         year = 2021,
        month = jan,
       volume = {906},
       number = {1},
          eid = {51},
        pages = {51},
          doi = {10.3847/1538-4357/abc886},
archivePrefix = {arXiv},
       eprint = {2011.03681},
 primaryClass = {astro-ph.HE},
       adsurl = {https://ui.adsabs.harvard.edu/abs/2021ApJ...906...51X}
}

@ARTICLE{Lobanov1998A&A330,
       author = {{Lobanov}, A.~P.},
        title = "{Ultracompact jets in active galactic nuclei}",
      journal = {\aap},
         year = 1998,
        month = feb,
       volume = {330},
        pages = {79-89},
          doi = {10.48550/arXiv.astro-ph/9712132},
archivePrefix = {arXiv},
       eprint = {astro-ph/9712132},
 primaryClass = {astro-ph},
       adsurl = {https://ui.adsabs.harvard.edu/abs/1998A&A...330...79L}
}

@ARTICLE{Xiao2022ApJ_1,
       author = {{Xiao}, Hubing and {Ouyang}, Zhihao and {Zhang}, Lixia and {Fu}, Liping and {Zhang}, Shaohua and {Zeng}, Xiangtao and {Fan}, Junhui},
        title = "{The Relativistic Jet and Central Engine of Fermi Blazars}",
      journal = {\apj},
         year = 2022,
        month = jan,
       volume = {925},
       number = {1},
          eid = {40},
        pages = {40},
          doi = {10.3847/1538-4357/ac36da},
archivePrefix = {arXiv},
       eprint = {2111.02082},
 primaryClass = {astro-ph.HE},
       adsurl = {https://ui.adsabs.harvard.edu/abs/2022ApJ...925...40X}
}

@ARTICLE{Yang2022ApJS,
       author = {{Yang}, J.~H. and {Fan}, J.~H. and {Liu}, Y. and {Tuo}, M.~X. and {Pei}, Z.~Y. and {Yang}, W.~X. and {Yuan}, Y.~H. and {He}, S.~L. and {Wang}, S.~H. and {Wang}, X.~C. and {Chen}, X.~J. and {Qu}, X.~H. and {Cao}, Q. and {Tao}, Q.~Y. and {Zhang}, Y.~L. and {Liu}, C.~Q. and {Nie}, J.~J. and {Liu}, L.~F. and {Jiang}, D.~K. and {Jiang}, A.~N. and {Liu}, B. and {Yang}, R.~S.},
        title = "{The Spectral Energy Distributions for 4FGL Blazars}",
      journal = {\apjs},
         year = 2022,
        month = sep,
       volume = {262},
       number = {1},
          eid = {18},
        pages = {18},
          doi = {10.3847/1538-4365/ac7deb},
       adsurl = {https://ui.adsabs.harvard.edu/abs/2022ApJS..262...18Y}
}

@ARTICLE{Xiao2022MNRAS,
       author = {{Xiao}, H.~B. and {Zhu}, J.~T. and {Fan}, J.~H. and {Pei}, Z.~Y. and {Luo}, Z.~J. and {Zhang}, S.~H.},
        title = "{The jet apparent motion and central engine study of Fermi blazars}",
      journal = {\mnras},
         year = 2022,
        month = dec,
       volume = {517},
       number = {3},
        pages = {4202-4212},
          doi = {10.1093/mnras/stac2959},
       adsurl = {https://ui.adsabs.harvard.edu/abs/2022MNRAS.517.4202X}
}

@ARTICLE{Xiao2022ApJ_2,
       author = {{Xiao}, Hubing and {Fan}, Junhui and {Ouyang}, Zhihao and {Hu}, Liangjun and {Chen}, Guohai and {Fu}, Liping and {Zhang}, Shaohua},
        title = "{An Extensive Study of Blazar Broad Emission Line: Changing-look Blazars and the Baldwin Effect}",
      journal = {\apj},
         year = 2022,
        month = sep,
       volume = {936},
       number = {2},
          eid = {146},
        pages = {146},
          doi = {10.3847/1538-4357/ac887f},
archivePrefix = {arXiv},
       eprint = {2208.10104},
 primaryClass = {astro-ph.HE},
       adsurl = {https://ui.adsabs.harvard.edu/abs/2022ApJ...936..146X}
}

@ARTICLE{Paliya2021,
       author = {{Paliya}, Vaidehi S. and {Dom{\'\i}nguez}, A. and {Ajello}, M. and {Olmo-Garc{\'\i}a}, A. and {Hartmann}, D.},
        title = "{The Central Engines of Fermi Blazars}",
      journal = {\apjs},
         year = 2021,
        month = apr,
       volume = {253},
       number = {2},
          eid = {46},
        pages = {46},
          doi = {10.3847/1538-4365/abe135},
archivePrefix = {arXiv},
       eprint = {2101.10849},
 primaryClass = {astro-ph.HE},
       adsurl = {https://ui.adsabs.harvard.edu/abs/2021ApJS..253...46P}
}

@ARTICLE{Zhang2020,
       author = {{Zhang}, Lixia and {Chen}, Sina and {Xiao}, Hubing and {Cai}, Jinting and {Fan}, Junhui},
        title = "{Doppler Factor Estimation for Fermi Blazars}",
      journal = {\apj},
         year = 2020,
        month = jul,
       volume = {897},
       number = {1},
          eid = {10},
        pages = {10},
          doi = {10.3847/1538-4357/ab9180},
       adsurl = {https://ui.adsabs.harvard.edu/abs/2020ApJ...897...10Z}
}

@ARTICLE{Xiao2019SCPMA,
       author = {{Xiao}, HuBing and {Fan}, JunHui and {Yang}, JiangHe and {Liu}, Yi and
         {Yuan}, YuHai and {Tao}, Jun and {Costantin}, Denise and
         {Zhang}, YuTao and {Pei}, ZhiYuan and {Zhang}, LiXia and {Yang}, WenXin},
        title = "{Comparison between Fermi detected and non-Fermi detected superluminal sources}",
      journal = {Science China Physics, Mechanics, and Astronomy},
         year = 2019,
        month = jul,
       volume = {62},
       number = {12},
          eid = {129811},
        pages = {129811},
          doi = {10.1007/s11433-018-9371-x},
archivePrefix = {arXiv},
       eprint = {1902.10764},
 primaryClass = {astro-ph.HE},
       adsurl = {https://ui.adsabs.harvard.edu/abs/2019SCPMA..62l9811X}
}

@ARTICLE{Marscher1985ApJ298,
       author = {{Marscher}, A.~P. and {Gear}, W.~K.},
        title = "{Models for high-frequency radio outbursts in extragalactic sources, with application to the early 1983 millimeter-to-infrared flare of 3C 273.}",
      journal = {\apj},
         year = 1985,
        month = nov,
       volume = {298},
        pages = {114-127},
          doi = {10.1086/163592},
       adsurl = {https://ui.adsabs.harvard.edu/abs/1985ApJ...298..114M}
}

@ARTICLE{Ghisellini2014,
       author = {{Ghisellini}, G. and {Tavecchio}, F. and {Maraschi}, L. and
         {Celotti}, A. and {Sbarrato}, T.},
        title = "{The power of relativistic jets is larger than the luminosity of their accretion disks}",
      journal = {\nat},
         year = 2014,
        month = nov,
       volume = {515},
       number = {7527},
        pages = {376-378},
          doi = {10.1038/nature13856},
archivePrefix = {arXiv},
       eprint = {1411.5368},
 primaryClass = {astro-ph.HE},
       adsurl = {https://ui.adsabs.harvard.edu/abs/2014Natur.515..376G}
}

@ARTICLE{Xiao2025ApJ991,
       author = {{Xiao}, Hubing and {Cao}, Haitao and {Xue}, Rui and {Ouyang}, Zhihao and {Zhang}, Shaohua and {Chen}, Junping and {Luo}, Zhijian and {Yang}, Jianghe and {Fan}, Junhui},
        title = "{The Averaged Broadband Spectral Energy Distribution Study of Fermi Bright BL Lac Objects}",
      journal = {\apj},
         year = 2025,
        month = sep,
       volume = {991},
       number = {1},
          eid = {78},
        pages = {78},
          doi = {10.3847/1538-4357/adf7ad},
archivePrefix = {arXiv},
       eprint = {2509.11820},
 primaryClass = {astro-ph.HE},
       adsurl = {https://ui.adsabs.harvard.edu/abs/2025ApJ...991...78X}
}

@ARTICLE{Xiong2014,
       author = {{Xiong}, D.~R. and {Zhang}, X.},
        title = "{Intrinsic {\ensuremath{\gamma}}-ray luminosity, black hole mass, jet and accretion in Fermi blazars}",
      journal = {\mnras},
         year = 2014,
        month = jul,
       volume = {441},
       number = {4},
        pages = {3375-3395},
          doi = {10.1093/mnras/stu755},
archivePrefix = {arXiv},
       eprint = {1404.3556},
 primaryClass = {astro-ph.HE},
       adsurl = {https://ui.adsabs.harvard.edu/abs/2014MNRAS.441.3375X}
}

@ARTICLE{Abdollahi2022ApJS260,
       author = {{Abdollahi}, S. and {Acero}, F. and {Baldini}, L. and {Ballet}, J. and {Bastieri}, D. and {Bellazzini}, R. and {Berenji}, B. and {Berretta}, A. and {Bissaldi}, E. and {Blandford}, R.~D. and {Bloom}, E. and {Bonino}, R. and {Brill}, A. and {Britto}, R.~J. and {Bruel}, P. and {Burnett}, T.~H. and {Buson}, S. and {Cameron}, R.~A. and {Caputo}, R. and {Caraveo}, P.~A. and {Castro}, D. and {Chaty}, S. and {Cheung}, C.~C. and {Chiaro}, G. and {Cibrario}, N. and {Ciprini}, S. and {Coronado-Bl{\'a}zquez}, J. and {Crnogorcevic}, M. and {Cutini}, S. and {D'Ammando}, F. and {De Gaetano}, S. and {Digel}, S.~W. and {Di Lalla}, N. and {Dirirsa}, F. and {Di Venere}, L. and {Dom{\'\i}nguez}, A. and {Fallah Ramazani}, V. and {Fegan}, S.~J. and {Ferrara}, E.~C. and {Fiori}, A. and {Fleischhack}, H. and {Franckowiak}, A. and {Fukazawa}, Y. and {Funk}, S. and {Fusco}, P. and {Galanti}, G. and {Gammaldi}, V. and {Gargano}, F. and {Garrappa}, S. and {Gasparrini}, D. and {Giacchino}, F. and {Giglietto}, N. and {Giordano}, F. and {Giroletti}, M. and {Glanzman}, T. and {Green}, D. and {Grenier}, I.~A. and {Grondin}, M. -H. and {Guillemot}, L. and {Guiriec}, S. and {Gustafsson}, M. and {Harding}, A.~K. and {Hays}, E. and {Hewitt}, J.~W. and {Horan}, D. and {Hou}, X. and {J{\'o}hannesson}, G. and {Karwin}, C. and {Kayanoki}, T. and {Kerr}, M. and {Kuss}, M. and {Landriu}, D. and {Larsson}, S. and {Latronico}, L. and {Lemoine-Goumard}, M. and {Li}, J. and {Liodakis}, I. and {Longo}, F. and {Loparco}, F. and {Lott}, B. and {Lubrano}, P. and {Maldera}, S. and {Malyshev}, D. and {Manfreda}, A. and {Mart{\'\i}-Devesa}, G. and {Mazziotta}, M.~N. and {Mereu}, I. and {Meyer}, M. and {Michelson}, P.~F. and {Mirabal}, N. and {Mitthumsiri}, W. and {Mizuno}, T. and {Moiseev}, A.~A. and {Monzani}, M.~E. and {Morselli}, A. and {Moskalenko}, I.~V. and {Negro}, M. and {Nuss}, E. and {Omodei}, N. and {Orienti}, M. and {Orlando}, E. and {Paneque}, D. and {Pei}, Z. and {Perkins}, J.~S. and {Persic}, M. and {Pesce-Rollins}, M. and {Petrosian}, V. and {Pillera}, R. and {Poon}, H. and {Porter}, T.~A. and {Principe}, G. and {Rain{\`o}}, S. and {Rando}, R. and {Rani}, B. and {Razzano}, M. and {Razzaque}, S. and {Reimer}, A. and {Reimer}, O. and {Reposeur}, T. and {S{\'a}nchez-Conde}, M. and {Saz Parkinson}, P.~M. and {Scotton}, L. and {Serini}, D. and {Sgr{\`o}}, C. and {Siskind}, E.~J. and {Smith}, D.~A. and {Spandre}, G. and {Spinelli}, P. and {Sueoka}, K. and {Suson}, D.~J. and {Tajima}, H. and {Tak}, D. and {Thayer}, J.~B. and {Thompson}, D.~J. and {Torres}, D.~F. and {Troja}, E. and {Valverde}, J. and {Wood}, K. and {Zaharijas}, G.},
        title = "{Incremental Fermi Large Area Telescope Fourth Source Catalog}",
      journal = {\apjs},
         year = 2022,
        month = jun,
       volume = {260},
       number = {2},
          eid = {53},
        pages = {53},
          doi = {10.3847/1538-4365/ac6751},
archivePrefix = {arXiv},
       eprint = {2201.11184},
 primaryClass = {astro-ph.HE},
       adsurl = {https://ui.adsabs.harvard.edu/abs/2022ApJS..260...53A}
}

@ARTICLE{Zhang2012ApJ752,
       author = {{Zhang}, Jin and {Liang}, En-Wei and {Zhang}, Shuang-Nan and {Bai}, J.~M.},
        title = "{Radiation Mechanisms and Physical Properties of GeV-TeV BL Lac Objects}",
      journal = {\apj},
         year = 2012,
        month = jun,
       volume = {752},
       number = {2},
          eid = {157},
        pages = {157},
          doi = {10.1088/0004-637X/752/2/157},
archivePrefix = {arXiv},
       eprint = {1108.0607},
 primaryClass = {astro-ph.HE},
       adsurl = {https://ui.adsabs.harvard.edu/abs/2012ApJ...752..157Z}
}

@ARTICLE{Neronov2010Sci328,
       author = {{Neronov}, Andrii and {Vovk}, Ievgen},
        title = "{Evidence for Strong Extragalactic Magnetic Fields from Fermi Observations of TeV Blazars}",
      journal = {Science},
         year = 2010,
        month = apr,
       volume = {328},
       number = {5974},
        pages = {73},
          doi = {10.1126/science.1184192},
archivePrefix = {arXiv},
       eprint = {1006.3504},
 primaryClass = {astro-ph.HE},
       adsurl = {https://ui.adsabs.harvard.edu/abs/2010Sci...328...73N}
}

@ARTICLE{Opher2009Natur462,
       author = {{Opher}, M. and {Bibi}, F. Alouani and {Toth}, G. and {Richardson}, J.~D. and {Izmodenov}, V.~V. and {Gombosi}, T.~I.},
        title = "{A strong, highly-tilted interstellar magnetic field near the Solar System}",
      journal = {\nat},
         year = 2009,
        month = dec,
       volume = {462},
       number = {7276},
        pages = {1036-1038},
          doi = {10.1038/nature08567},
       adsurl = {https://ui.adsabs.harvard.edu/abs/2009Natur.462.1036O}
}

@ARTICLE{Dolginov1988PhR162,
       author = {{Dolginov}, A.~Z.},
        title = "{Magnetic field generation in celestial bodies.}",
      journal = {\physrep},
         year = 1988,
        month = jan,
       volume = {162},
       number = {6},
        pages = {337-415},
          doi = {10.1016/0370-1573(88)90147-0},
       adsurl = {https://ui.adsabs.harvard.edu/abs/1988PhR...162..337D}
}

@ARTICLE{Padovani2017NatAs1,
       author = {{Padovani}, Paolo},
        title = "{On the two main classes of active galactic nuclei}",
      journal = {Nature Astronomy},
         year = 2017,
        month = aug,
       volume = {1},
          eid = {0194},
        pages = {0194},
          doi = {10.1038/s41550-017-0194},
archivePrefix = {arXiv},
       eprint = {1707.08069},
 primaryClass = {astro-ph.GA},
       adsurl = {https://ui.adsabs.harvard.edu/abs/2017NatAs...1E.194P}
}

@ARTICLE{Kim2022MNRAS510,
       author = {{Kim}, Sang-Hyun and {Lee}, Sang-Sung and {Lee}, Jee Won and {Hodgson}, Jeffrey A. and {Kang}, Sincheol and {Algaba}, Juan-Carlos and {Kim}, Jae-Young and {Hodges}, Mark and {Agudo}, Ivan and {Fuentes}, Antonio and {Escudero}, Juan and {Myserlis}, Ioannis and {Traianou}, Efthalia and {L{\"a}hteenm{\"a}ki}, Anne and {Tornikoski}, Merja and {Tammi}, Joni and {Ramakrishnan}, Venkatessh and {J{\"a}rvel{\"a}}, Emilia},
        title = "{Magnetic field strengths of the synchrotron self-absorption region in the jet of CTA 102 during radio flares}",
      journal = {\mnras},
         year = 2022,
        month = feb,
       volume = {510},
       number = {1},
        pages = {815-833},
          doi = {10.1093/mnras/stab3473},
archivePrefix = {arXiv},
       eprint = {2111.14025},
 primaryClass = {astro-ph.HE},
       adsurl = {https://ui.adsabs.harvard.edu/abs/2022MNRAS.510..815K}
}

@ARTICLE{Vlahakis2004ApJ605,
       author = {{Vlahakis}, Nektarios and {K{\"o}nigl}, Arieh},
        title = "{Magnetic Driving of Relativistic Outflows in Active Galactic Nuclei. I. Interpretation of Parsec-Scale Accelerations}",
      journal = {\apj},
         year = 2004,
        month = apr,
       volume = {605},
       number = {2},
        pages = {656-661},
          doi = {10.1086/382670},
archivePrefix = {arXiv},
       eprint = {astro-ph/0310747},
 primaryClass = {astro-ph},
       adsurl = {https://ui.adsabs.harvard.edu/abs/2004ApJ...605..656V}
}

@ARTICLE{Komissarov2007MNRAS380,
       author = {{Komissarov}, Serguei S. and {Barkov}, Maxim V. and {Vlahakis}, Nektarios and {K{\"o}nigl}, Arieh},
        title = "{Magnetic acceleration of relativistic active galactic nucleus jets}",
      journal = {\mnras},
         year = 2007,
        month = sep,
       volume = {380},
       number = {1},
        pages = {51-70},
          doi = {10.1111/j.1365-2966.2007.12050.x},
archivePrefix = {arXiv},
       eprint = {astro-ph/0703146},
 primaryClass = {astro-ph},
       adsurl = {https://ui.adsabs.harvard.edu/abs/2007MNRAS.380...51K}
}

@ARTICLE{Lee2016ApJ826,
       author = {{Lee}, Sang-Sung and {Lobanov}, Andrei P. and {Krichbaum}, Thomas P. and {Zensus}, J. Anton},
        title = "{Acceleration of Compact Radio Jets on Sub-parsec Scales}",
      journal = {\apj},
         year = 2016,
        month = aug,
       volume = {826},
       number = {2},
          eid = {135},
        pages = {135},
          doi = {10.3847/0004-637X/826/2/135},
archivePrefix = {arXiv},
       eprint = {1604.02207},
 primaryClass = {astro-ph.GA},
       adsurl = {https://ui.adsabs.harvard.edu/abs/2016ApJ...826..135L}
}

@ARTICLE{Konigl1981ApJ243,
       author = {{Konigl}, A.},
        title = "{Relativistic jets as X-ray and gamma-ray sources.}",
      journal = {\apj},
         year = 1981,
        month = feb,
       volume = {243},
        pages = {700-709},
          doi = {10.1086/158638},
       adsurl = {https://ui.adsabs.harvard.edu/abs/1981ApJ...243..700K}
}

@ARTICLE{Zdziarski2015MNRAS451,
       author = {{Zdziarski}, Andrzej A. and {Sikora}, Marek and {Pjanka}, Patryk and {Tchekhovskoy}, Alexander},
        title = "{Core shifts, magnetic fields and magnetization of extragalactic jets}",
      journal = {\mnras},
         year = 2015,
        month = jul,
       volume = {451},
       number = {1},
        pages = {927-935},
          doi = {10.1093/mnras/stv986},
archivePrefix = {arXiv},
       eprint = {1410.7310},
 primaryClass = {astro-ph.HE},
       adsurl = {https://ui.adsabs.harvard.edu/abs/2015MNRAS.451..927Z}
}

@ARTICLE{Marscher1988ApJ334,
       author = {{Marscher}, Alan P.},
        title = "{Contemporaneous X-Ray and VLBI Radio Observations of the Quasar NRAO 140}",
      journal = {\apj},
         year = 1988,
        month = nov,
       volume = {334},
        pages = {552},
          doi = {10.1086/166859},
       adsurl = {https://ui.adsabs.harvard.edu/abs/1988ApJ...334..552M}
}

@ARTICLE{Planck2016A&A594,
       author = {{Planck Collaboration} and {Ade}, P.~A.~R. and {Aghanim}, N. and {Arnaud}, M. and {Arroja}, F. and {Ashdown}, M. and {Aumont}, J. and {Baccigalupi}, C. and {Ballardini}, M. and {Banday}, A.~J. and {Barreiro}, R.~B. and {Bartolo}, N. and {Battaner}, E. and {Benabed}, K. and {Beno{\^\i}t}, A. and {Benoit-L{\'e}vy}, A. and {Bernard}, J. -P. and {Bersanelli}, M. and {Bielewicz}, P. and {Bock}, J.~J. and {Bonaldi}, A. and {Bonavera}, L. and {Bond}, J.~R. and {Borrill}, J. and {Bouchet}, F.~R. and {Bucher}, M. and {Burigana}, C. and {Butler}, R.~C. and {Calabrese}, E. and {Cardoso}, J. -F. and {Catalano}, A. and {Chamballu}, A. and {Chiang}, H.~C. and {Chluba}, J. and {Christensen}, P.~R. and {Church}, S. and {Clements}, D.~L. and {Colombi}, S. and {Colombo}, L.~P.~L. and {Combet}, C. and {Couchot}, F. and {Coulais}, A. and {Crill}, B.~P. and {Curto}, A. and {Cuttaia}, F. and {Danese}, L. and {Davies}, R.~D. and {Davis}, R.~J. and {de Bernardis}, P. and {de Rosa}, A. and {de Zotti}, G. and {Delabrouille}, J. and {D{\'e}sert}, F. -X. and {Diego}, J.~M. and {Dolag}, K. and {Dole}, H. and {Donzelli}, S. and {Dor{\'e}}, O. and {Douspis}, M. and {Ducout}, A. and {Dupac}, X. and {Efstathiou}, G. and {Elsner}, F. and {En{\ss}lin}, T.~A. and {Eriksen}, H.~K. and {Fergusson}, J. and {Finelli}, F. and {Florido}, E. and {Forni}, O. and {Frailis}, M. and {Fraisse}, A.~A. and {Franceschi}, E. and {Frejsel}, A. and {Galeotta}, S. and {Galli}, S. and {Ganga}, K. and {Giard}, M. and {Giraud-H{\'e}raud}, Y. and {Gjerl{\o}w}, E. and {Gonz{\'a}lez-Nuevo}, J. and {G{\'o}rski}, K.~M. and {Gratton}, S. and {Gregorio}, A. and {Gruppuso}, A. and {Gudmundsson}, J.~E. and {Hansen}, F.~K. and {Hanson}, D. and {Harrison}, D.~L. and {Helou}, G. and {Henrot-Versill{\'e}}, S. and {Hern{\'a}ndez-Monteagudo}, C. and {Herranz}, D. and {Hildebrandt}, S.~R. and {Hivon}, E. and {Hobson}, M. and {Holmes}, W.~A. and {Hornstrup}, A. and {Hovest}, W. and {Huffenberger}, K.~M. and {Hurier}, G. and {Jaffe}, A.~H. and {Jaffe}, T.~R. and {Jones}, W.~C. and {Juvela}, M. and {Keih{\"a}nen}, E. and {Keskitalo}, R. and {Kim}, J. and {Kisner}, T.~S. and {Knoche}, J. and {Kunz}, M. and {Kurki-Suonio}, H. and {Lagache}, G. and {L{\"a}hteenm{\"a}ki}, A. and {Lamarre}, J. -M. and {Lasenby}, A. and {Lattanzi}, M. and {Lawrence}, C.~R. and {Leahy}, J.~P. and {Leonardi}, R. and {Lesgourgues}, J. and {Levrier}, F. and {Liguori}, M. and {Lilje}, P.~B. and {Linden-V{\o}rnle}, M. and {L{\'o}pez-Caniego}, M. and {Lubin}, P.~M. and {Mac{\'\i}as-P{\'e}rez}, J.~F. and {Maggio}, G. and {Maino}, D. and {Mandolesi}, N. and {Mangilli}, A. and {Maris}, M. and {Martin}, P.~G. and {Mart{\'\i}nez-Gonz{\'a}lez}, E. and {Masi}, S. and {Matarrese}, S. and {McGehee}, P. and {Meinhold}, P.~R. and {Melchiorri}, A. and {Mendes}, L. and {Mennella}, A. and {Migliaccio}, M. and {Mitra}, S. and {Miville-Desch{\^e}nes}, M. -A. and {Molinari}, D. and {Moneti}, A. and {Montier}, L. and {Morgante}, G. and {Mortlock}, D. and {Moss}, A. and {Munshi}, D. and {Murphy}, J.~A. and {Naselsky}, P. and {Nati}, F. and {Natoli}, P. and {Netterfield}, C.~B. and {N{\o}rgaard-Nielsen}, H.~U. and {Noviello}, F. and {Novikov}, D. and {Novikov}, I. and {Oppermann}, N. and {Oxborrow}, C.~A. and {Paci}, F. and {Pagano}, L. and {Pajot}, F. and {Paoletti}, D. and {Pasian}, F. and {Patanchon}, G. and {Perdereau}, O. and {Perotto}, L. and {Perrotta}, F. and {Pettorino}, V. and {Piacentini}, F. and {Piat}, M. and {Pierpaoli}, E. and {Pietrobon}, D. and {Plaszczynski}, S. and {Pointecouteau}, E. and {Polenta}, G. and {Popa}, L. and {Pratt}, G.~W. and {Pr{\'e}zeau}, G. and {Prunet}, S. and {Puget}, J. -L. and {Rachen}, J.~P. and {Rebolo}, R. and {Reinecke}, M. and {Remazeilles}, M. and {Renault}, C. and {Renzi}, A. and {Ristorcelli}, I. and {Rocha}, G. and {Rosset}, C. and {Rossetti}, M. and {Roudier}, G. and {Rubi{\~n}o-Mart{\'\i}n}, J.~A. and {Ruiz-Granados}, B. and {Rusholme}, B. and {Sandri}, M. and {Santos}, D. and {Savelainen}, M. and {Savini}, G. and {Scott}, D. and {Seiffert}, M.~D. and {Shellard}, E.~P.~S. and {Shiraishi}, M. and {Spencer}, L.~D. and {Stolyarov}, V. and {Stompor}, R. and {Sudiwala}, R. and {Sunyaev}, R. and {Sutton}, D. and {Suur-Uski}, A. -S. and {Sygnet}, J. -F. and {Tauber}, J.~A. and {Terenzi}, L. and {Toffolatti}, L. and {Tomasi}, M. and {Tristram}, M. and {Tucci}, M. and {Tuovinen}, J. and {Umana}, G. and {Valenziano}, L. and {Valiviita}, J. and {Van Tent}, B. and {Vielva}, P. and {Villa}, F. and {Wade}, L.~A. and {Wandelt}, B.~D. and {Wehus}, I.~K. and {Yvon}, D. and {Zacchei}, A. and {Zonca}, A.},
        title = "{Planck 2015 results. XIX. Constraints on primordial magnetic fields}",
      journal = {\aap},
         year = 2016,
        month = sep,
       volume = {594},
          eid = {A19},
        pages = {A19},
          doi = {10.1051/0004-6361/201525821},
archivePrefix = {arXiv},
       eprint = {1502.01594},
 primaryClass = {astro-ph.CO},
       adsurl = {https://ui.adsabs.harvard.edu/abs/2016A&A...594A..19P}
}

@ARTICLE{Chen2018,
       author = {{Chen}, Liang},
        title = "{On the Jet Properties of {\ensuremath{\gamma}}-Ray-loud Active Galactic Nuclei}",
      journal = {\apjs},
         year = 2018,
        month = apr,
       volume = {235},
       number = {2},
          eid = {39},
        pages = {39},
          doi = {10.3847/1538-4365/aab8fb},
archivePrefix = {arXiv},
       eprint = {1803.05715},
 primaryClass = {astro-ph.HE},
       adsurl = {https://ui.adsabs.harvard.edu/abs/2018ApJS..235...39C}
}

@ARTICLE{Fan2023ApJS,
       author = {{Fan}, Junhui and {Xiao}, Hubing and {Yang}, Wenxin and {Zhang}, Lixia and {Strigachev}, Anton A. and {Bachev}, Rumen S. and {Yang}, Jianghe},
        title = "{Characterizing the Emission Region Properties of Blazars}",
      journal = {\apjs},
         year = 2023,
        month = sep,
       volume = {268},
       number = {1},
          eid = {23},
        pages = {23},
          doi = {10.3847/1538-4365/ace7c8},
archivePrefix = {arXiv},
       eprint = {2307.07163},
 primaryClass = {astro-ph.HE},
       adsurl = {https://ui.adsabs.harvard.edu/abs/2023ApJS..268...23F}
}

@ARTICLE{Ghisellini2009MNRAS397,
       author = {{Ghisellini}, G. and {Tavecchio}, F.},
        title = "{Canonical high-power blazars}",
      journal = {\mnras},
         year = 2009,
        month = aug,
       volume = {397},
       number = {2},
        pages = {985-1002},
          doi = {10.1111/j.1365-2966.2009.15007.x},
archivePrefix = {arXiv},
       eprint = {0902.0793},
 primaryClass = {astro-ph.CO},
       adsurl = {https://ui.adsabs.harvard.edu/abs/2009MNRAS.397..985G}
}

@ARTICLE{Blandford1977,
       author = {{Blandford}, R.~D. and {Znajek}, R.~L.},
        title = "{Electromagnetic extraction of energy from Kerr black holes.}",
      journal = {\mnras},
         year = 1977,
        month = may,
       volume = {179},
        pages = {433-456},
          doi = {10.1093/mnras/179.3.433},
       adsurl = {https://ui.adsabs.harvard.edu/abs/1977MNRAS.179..433B}
}

@ARTICLE{Blandford1982,
       author = {{Blandford}, R.~D. and {Payne}, D.~G.},
        title = "{Hydromagnetic flows from accretion disks and the production of radio jets.}",
      journal = {\mnras},
         year = 1982,
        month = jun,
       volume = {199},
        pages = {883-903},
          doi = {10.1093/mnras/199.4.883},
       adsurl = {https://ui.adsabs.harvard.edu/abs/1982MNRAS.199..883B}
}

@ARTICLE{Turler2000AA361,
       author = {{T{\"u}rler}, M. and {Courvoisier}, T.~J. -L. and {Paltani}, S.},
        title = "{Modelling 20 years of synchrotron flaring in the jet of 3C 273}",
      journal = {\aap},
         year = 2000,
        month = sep,
       volume = {361},
        pages = {850-862},
          doi = {10.48550/arXiv.astro-ph/0008480},
archivePrefix = {arXiv},
       eprint = {astro-ph/0008480},
 primaryClass = {astro-ph},
       adsurl = {https://ui.adsabs.harvard.edu/abs/2000A&A...361..850T}
}

@ARTICLE{Blandford1979ApJ232,
       author = {{Blandford}, R.~D. and {K{\"o}nigl}, A.},
        title = "{Relativistic jets as compact radio sources.}",
      journal = {\apj},
         year = 1979,
        month = aug,
       volume = {232},
        pages = {34-48},
          doi = {10.1086/157262},
       adsurl = {https://ui.adsabs.harvard.edu/abs/1979ApJ...232...34B}
}

@ARTICLE{Hovatta2009,
       author = {{Hovatta}, T. and {Valtaoja}, E. and {Tornikoski}, M. and {L{\"a}hteenm{\"a}ki}, A.},
        title = "{Doppler factors, Lorentz factors and viewing angles for quasars, BL Lacertae objects and radio galaxies}",
      journal = {\aap},
         year = 2009,
        month = feb,
       volume = {494},
       number = {2},
        pages = {527-537},
          doi = {10.1051/0004-6361:200811150},
archivePrefix = {arXiv},
       eprint = {0811.4278},
 primaryClass = {astro-ph},
       adsurl = {https://ui.adsabs.harvard.edu/abs/2009A&A...494..527H}
}

@ARTICLE{Liodakis2018,
       author = {{Liodakis}, Ioannis and {Hovatta}, Talvikki and {Huppenkothen}, Daniela and {Kiehlmann}, Sebastian and {Max-Moerbeck}, Walter and {Readhead}, Anthony C.~S.},
        title = "{Constraining the Limiting Brightness Temperature and Doppler Factors for the Largest Sample of Radio-bright Blazars}",
      journal = {\apj},
         year = 2018,
        month = oct,
       volume = {866},
       number = {2},
          eid = {137},
        pages = {137},
          doi = {10.3847/1538-4357/aae2b7},
archivePrefix = {arXiv},
       eprint = {1809.08249},
 primaryClass = {astro-ph.HE},
       adsurl = {https://ui.adsabs.harvard.edu/abs/2018ApJ...866..137L}
}

@ARTICLE{Hovatta2019AA,
       author = {{Hovatta}, T. and {O'Sullivan}, S. and {Mart{\'\i}-Vidal}, I. and {Savolainen}, T. and {Tchekhovskoy}, A.},
        title = "{Magnetic field at a jet base: extreme Faraday rotation in 3C 273 revealed by ALMA}",
      journal = {\aap},
         year = 2019,
        month = mar,
       volume = {623},
          eid = {A111},
        pages = {A111},
          doi = {10.1051/0004-6361/201832587},
archivePrefix = {arXiv},
       eprint = {1803.09982},
 primaryClass = {astro-ph.GA},
       adsurl = {https://ui.adsabs.harvard.edu/abs/2019A&A...623A.111H}
}

@ARTICLE{Ouyang2025ApJ980,
       author = {{Ouyang}, Zhihao and {Xiao}, Hubing and {Manganaro}, Marina and {Xie}, Shangchun and {Wu}, Jingyu and {Chen}, Jianzhen and {Xue}, Rui and {Wang}, Gege and {Zhang}, Shaohua and {Fan}, Junhui},
        title = "{Revisiting the Flaring Activity in Early 2015 of BL Lacertae Object S5 0716+714}",
      journal = {\apj},
         year = 2025,
        month = feb,
       volume = {980},
       number = {1},
          eid = {19},
        pages = {19},
          doi = {10.3847/1538-4357/ada3bc},
archivePrefix = {arXiv},
       eprint = {2412.20842},
 primaryClass = {astro-ph.HE},
       adsurl = {https://ui.adsabs.harvard.edu/abs/2025ApJ...980...19O}
}

@ARTICLE{Fromm2011AA531,
       author = {{Fromm}, C.~M. and {Perucho}, M. and {Ros}, E. and {Savolainen}, T. and {Lobanov}, A.~P. and {Zensus}, J.~A. and {Aller}, M.~F. and {Aller}, H.~D. and {Gurwell}, M.~A. and {L{\"a}hteenm{\"a}ki}, A.},
        title = "{Catching the radio flare in CTA 102. I. Light curve analysis}",
      journal = {\aap},
         year = 2011,
        month = jul,
       volume = {531},
          eid = {A95},
        pages = {A95},
          doi = {10.1051/0004-6361/201116857},
archivePrefix = {arXiv},
       eprint = {1105.5024},
 primaryClass = {astro-ph.CO},
       adsurl = {https://ui.adsabs.harvard.edu/abs/2011A&A...531A..95F}
}

@ARTICLE{Marscher1983ApJ264,
       author = {{Marscher}, A.~P.},
        title = "{Accurate formula for the self-Compton X-ray flux density from a uniform, spherical, compact radio source.}",
      journal = {\apj},
         year = 1983,
        month = jan,
       volume = {264},
        pages = {296-297},
          doi = {10.1086/160597},
       adsurl = {https://ui.adsabs.harvard.edu/abs/1983ApJ...264..296M}
}

@ARTICLE{Rani2013,
       author = {{Rani}, B. and {Krichbaum}, T.~P. and {Fuhrmann}, L. and {B{\"o}ttcher}, M. and {Lott}, B. and {Aller}, H.~D. and {Aller}, M.~F. and {Angelakis}, E. and {Bach}, U. and {Bastieri}, D. and {Falcone}, A.~D. and {Fukazawa}, Y. and {Gabanyi}, K.~E. and {Gupta}, A.~C. and {Gurwell}, M. and {Itoh}, R. and {Kawabata}, K.~S. and {Krips}, M. and {L{\"a}hteenm{\"a}ki}, A.~A. and {Liu}, X. and {Marchili}, N. and {Max-Moerbeck}, W. and {Nestoras}, I. and {Nieppola}, E. and {Quintana-Lacaci}, G. and {Readhead}, A.~C.~S. and {Richards}, J.~L. and {Sasada}, M. and {Sievers}, A. and {Sokolovsky}, K. and {Stroh}, M. and {Tammi}, J. and {Tornikoski}, M. and {Uemura}, M. and {Ungerechts}, H. and {Urano}, T. and {Zensus}, J.~A.},
        title = "{Radio to gamma-ray variability study of blazar S5 0716+714}",
      journal = {\aap},
         year = 2013,
        month = apr,
       volume = {552},
          eid = {A11},
        pages = {A11},
          doi = {10.1051/0004-6361/201321058},
archivePrefix = {arXiv},
       eprint = {1301.7087},
 primaryClass = {astro-ph.HE},
       adsurl = {https://ui.adsabs.harvard.edu/abs/2013A&A...552A..11R}
}

\end{document}